\documentclass[aps,prb,twocolumn,floatfix] {revtex4-2}
\usepackage{graphicx,amsfonts}
\usepackage{bm,color}
\usepackage{multirow}
\usepackage{amssymb,amsmath,hyperref}
\usepackage{wrapfig}
\usepackage{lineno}
\usepackage{float}
\DeclareUnicodeCharacter{2212}{-}

\RequirePackage[normalem]{ulem} 
\RequirePackage{color}\definecolor{RED}{rgb}{1,0,0}\definecolor{BLUE}{rgb}{0,0,1} 
\providecommand{\DIFaddbegin}{} 
\providecommand{\DIFaddend}{} 
\providecommand{\DIFdelbegin}{} 
\providecommand{\DIFdelend}{} 
\providecommand{\DIFaddbeginFL}{} 
\providecommand{\DIFaddendFL}{} 
\providecommand{\DIFdelbeginFL}{} 
\providecommand{\DIFdelendFL}{} 
\newcommand{\DIFscaledelfig}{0.5}

\RequirePackage{settobox} 
\RequirePackage{letltxmacro} 
\newsavebox{\DIFdelgraphicsbox} 
\newlength{\DIFdelgraphicswidth} 
\newlength{\DIFdelgraphicsheight} 
\LetLtxMacro{\DIFOincludegraphics}{\includegraphics} 
\newcommand{\DIFaddincludegraphics}[2][]{{\color{blue}\fbox{\DIFOincludegraphics[#1]{#2}}}} 
\newcommand{\DIFdelincludegraphics}[2][]{
\sbox{\DIFdelgraphicsbox}{\DIFOincludegraphics[#1]{#2}}
\settoboxwidth{\DIFdelgraphicswidth}{\DIFdelgraphicsbox} 
\settoboxtotalheight{\DIFdelgraphicsheight}{\DIFdelgraphicsbox} 
\scalebox{\DIFscaledelfig}{
\parbox[b]{\DIFdelgraphicswidth}{\usebox{\DIFdelgraphicsbox}\\[-\baselineskip] \rule{\DIFdelgraphicswidth}{0em}}\llap{\resizebox{\DIFdelgraphicswidth}{\DIFdelgraphicsheight}{
\setlength{\unitlength}{\DIFdelgraphicswidth}
\begin{picture}(1,1)
\thicklines\linethickness{2pt} 
{\color[rgb]{1,0,0}\put(0,0){\framebox(1,1){}}}
{\color[rgb]{1,0,0}\put(0,0){\line( 1,1){1}}}
{\color[rgb]{1,0,0}\put(0,1){\line(1,-1){1}}}
\end{picture}
}\hspace*{3pt}}} 
} 
\LetLtxMacro{\DIFOaddbegin}{\DIFaddbegin} 
\LetLtxMacro{\DIFOaddend}{\DIFaddend} 
\LetLtxMacro{\DIFOdelbegin}{\DIFdelbegin} 
\LetLtxMacro{\DIFOdelend}{\DIFdelend} 
\DeclareRobustCommand{\DIFaddbegin}{\DIFOaddbegin \let\includegraphics\DIFaddincludegraphics} 
\DeclareRobustCommand{\DIFaddend}{\DIFOaddend \let\includegraphics\DIFOincludegraphics} 
\DeclareRobustCommand{\DIFdelbegin}{\DIFOdelbegin \let\includegraphics\DIFdelincludegraphics} 
\DeclareRobustCommand{\DIFdelend}{\DIFOaddend \let\includegraphics\DIFOincludegraphics} 
\LetLtxMacro{\DIFOaddbeginFL}{\DIFaddbeginFL} 
\LetLtxMacro{\DIFOaddendFL}{\DIFaddendFL} 
\LetLtxMacro{\DIFOdelbeginFL}{\DIFdelbeginFL} 
\LetLtxMacro{\DIFOdelendFL}{\DIFdelendFL} 
\DeclareRobustCommand{\DIFaddbeginFL}{\DIFOaddbeginFL \let\includegraphics\DIFaddincludegraphics} 
\DeclareRobustCommand{\DIFaddendFL}{\DIFOaddendFL \let\includegraphics\DIFOincludegraphics} 
\DeclareRobustCommand{\DIFdelbeginFL}{\DIFOdelbeginFL \let\includegraphics\DIFdelincludegraphics} 
\DeclareRobustCommand{\DIFdelendFL}{\DIFOaddendFL \let\includegraphics\DIFOincludegraphics} 

\usepackage{braket}

\begin{document}


\title{Heat measurement of quantum interference}

\author{Christoforus Dimas Satrya$^1$}
\email{christoforus.satrya@aalto.fi}

\author{Aleksandr S. Strelnikov$^1$}
\author{Luca Magazz\`u$^{1}$}
 \author{Yu-Cheng~Chang$^{1}$}
 \author{Rishabh Upadhyay$^{1,2}$}
 \author{Joonas T. Peltonen$^1$}
 \author{Bayan Karimi$^{1,3}$}

\author{Jukka P. Pekola$^1$}
\email{jukka.pekola@aalto.fi}

 \affiliation{$^1$ Pico group, QTF Centre of Excellence, Department of Applied Physics, Aalto University School of Science, P.O. Box 13500, 00076 Aalto, Finland}

\affiliation{$^2$ VTT Technical Research Centre of Finland Ltd, Tietotie 3, 02150 Espoo, Finland}

 \affiliation{$^3$ Pritzker School of Molecular Engineering, University of Chicago, Chicago IL 60637, USA}


\begin{abstract}
Coherence is a key property of quantum systems, and it plays a central role in the operation and performance of quantum heat engines and refrigerators. Despite its importance for the fundamental understanding in quantum thermodynamics and its technological implications, coherence effects in heat transport have not been observed previously. Here, we measure quantum features in the heat transfer between a qubit and a thermal bath. The system is formed of a driven flux qubit galvanically coupled to a $\lambda/4$ coplanar-waveguide resonator that is coupled to a heat reservoir. This thermal bath is a normal-metal mesoscopic resistor, whose temperature can be measured and controlled. We detect interference patterns in the heat current due to driving-induced coherence. In particular, resonance peaks in the heat transferred to the bath are found at driving frequencies which are integer fractions of the resonator frequency. A selection rule on the even/odd parity of the peaks holds at the qubit symmetry point. We present a theoretical model based on Floquet theory that captures the experimental results. The studied system provides a platform for studying the role of coherence in quantum thermodynamics. Our work opens the possibility to demonstrate a true quantum thermal machine where heat is measured directly.

  
\end{abstract}




\maketitle

\section*{Introduction}



Quantum heat engines and refrigerators have attracted considerable interest as they provide model systems for investigating quantum thermodynamics \cite{annurev-physchem-040513-103724,Alicki_1979,PhysRevE.76.031105,Binder2019}. Besides their fundamental relevance for understanding thermodynamics in the quantum regime \cite{Campbell2025,e15062100}, they also offer practical applications to quantum technology and information \cite{Aamir2025}. The key distinction between the working substance of a quantum thermal machine and a classical one lies in the ability to exist in a coherent superposition of states \cite{Uzdin2015,PhysRevLett.122.110601,PhysRevLett.125.166802}. However, whether quantum coherence is advantageous for the performance of a thermodynamic cycle is currently the topic of much active research \cite{Scully2003,PhysRevX.5.031044,Jaramillo2016,Scully2011,PhysRevA.99.062103,PhysRevLett.127.190604,PhysRevLett.119.170602,PhysRevE.93.062134}. An instance is the Otto cycle setup, where driving-induced coherence is found to reduce the efficiency and cooling power of quantum refrigerators \cite{PhysRevB.94.184503,Thomas2023}. The origin of the degradation in the performance, for non-adiabatic driving, is the constructive interference between the qubit states resulting in a finite steady-state population of the excited state. This phenomenon, known as Landau-Zener-St\"uckelberg-Majorana (LZSM) interference, occurs when a qubit \cite{Oliver2005,Sillanpaeae2006,Ivakhnenko2023,Grifoni1998,Silveri2017} or qubit-resonator system \cite{PhysRevLett.111.137002,PhysRevLett.123.240401,Chen2021,PhysRevLett.130.233602,probingstronglydrivenstrongly2025} is periodically driven across an avoided crossing inducing transitions between the states. Driving protocols that avoid creating such coherence and thus restore the performance of the Otto refrigerator to its optimum classical value are proposed in \cite{PhysRevB.100.085405, PhysRevB.100.035407}. 

In circuit quantum electrodynamics (c-QED) \cite{blais_2021_circuit,blais_2020_quantum}, a dissipative element such as a normal-metal resistor can be integrated and used as an engineered thermal bath. Its thermal properties and the experimental techniques for temperature manipulation and reading are well understood \cite{Giazotto2006,pekola_2021_icolloquiumi}, providing a tool to directly measure a steady-state heat current from a quantum system to the resistor. This has been experimentally demonstrated in the realization of thermal devices such as the quantum heat valve \cite{Ronzani2018} and diode \cite{Senior2020}. These devices consist of two heat baths that connect to a qubit via resonators, and the heat transport between the heat baths is controlled by the value of the qubit energy. In such experiments, transitions in the qubit are excited by thermal noise from the resistors and the heat current is driven by a temperature bias. Other applications of resistors integrated in superconducting circuits for thermal devices and sensors are given in \cite{MartinezPerez2015,PhysRevLett.125.237701,Giazotto2012,yoshioka_2023_active,partanen_2018_fluxtunable,tan_2017_quantumcircuit,Gunyho2024}. If the qubit is coupled
to a single heat bath, a steady-state heat current to the
bath can be induced by a periodic drive on the qubit~\cite{Carrega2016, Thomas2023}.



\begin{figure*}[t!]
\includegraphics{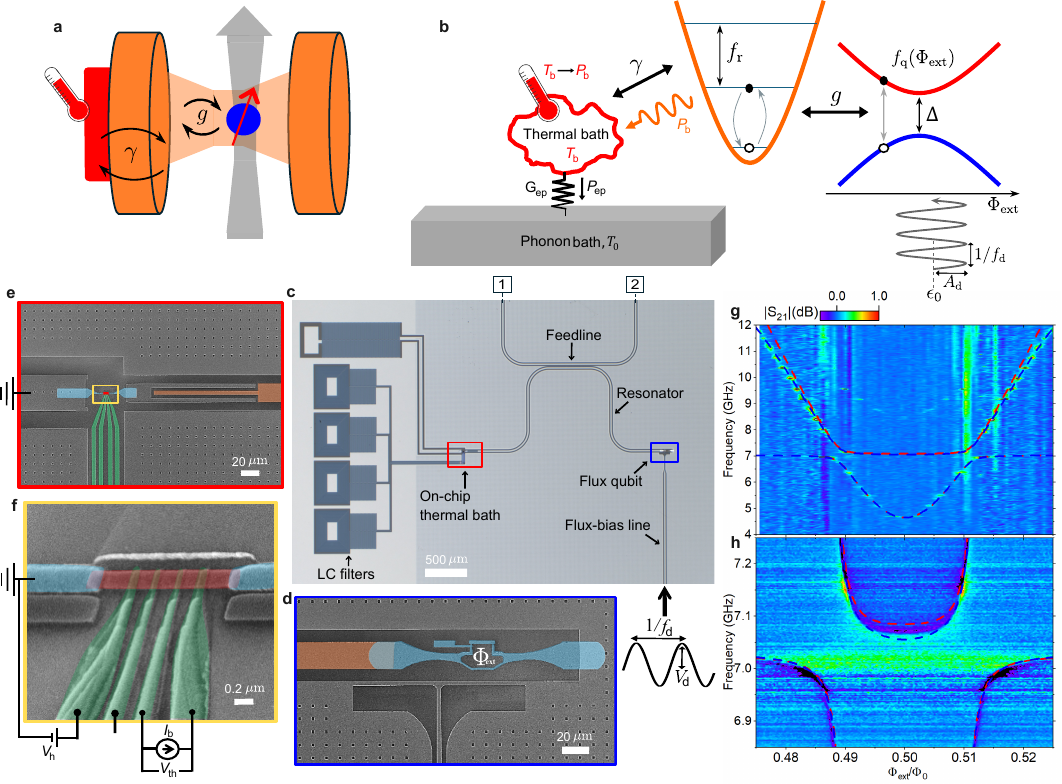}
 \caption{\textbf{The studied system and device}. \textbf{a}, Conceptual description of the studied system consisting of a driven qubit embedded in a cavity. The coupling $g$ is the energy exchange rate between qubit and cavity, while the coupling $\gamma$ is the energy exchange rate between the cavity and its environment bath. \textbf{b}, Schematic of a driven qubit-resonator coupled weakly to a thermal bath. The driven system transfers heat $P_{\text{b}}$ to the thermal bath raising its temperature $T_{\text{b}}$. The heat transferred to the bath is measured by observing the steady-state temperature difference between the bath and the substrate (phonon bath). \textbf{c}, Optical micrograph of the fabricated device. A $\lambda/4$ resonator is shunted by a flux qubit (inside blue rectangle). The open side of the resonator is weakly capacitively attached to an on-chip thermal bath made of a Cu film (inside red rectangle). The resonator is weakly attached to a feedline for RF scattering measurement. \textbf{d}, SEM image of the flux qubit made of Al film (blue) whose frequency $f_{\text{q}}$ can be tuned by an external magnetic flux $\Phi_{\text{ext}}$. The resonator made of Nb film (orange) is clean-contacted to the Al film. \textbf{e}, SEM image of the resonator (orange) capacitively coupled to the Cu film (inside yellow rectangle). \textbf{f}, SEM image of the Cu film (red) that is connected to four superconducting Al leads (green) with an insulating layer, forming NIS junctions for probing the electronic temperature $T_{\text{b}}$ in the Cu resistor. \textbf{g}, Two-tone spectroscopy showing the qubit-resonator spectrum. \textbf{h}, One-tone spectroscopy showing Rabi splittings where the qubit and resonator are at resonance.  \label{fig:1}}   
\end{figure*}

\begin{figure*}[t!]

\includegraphics{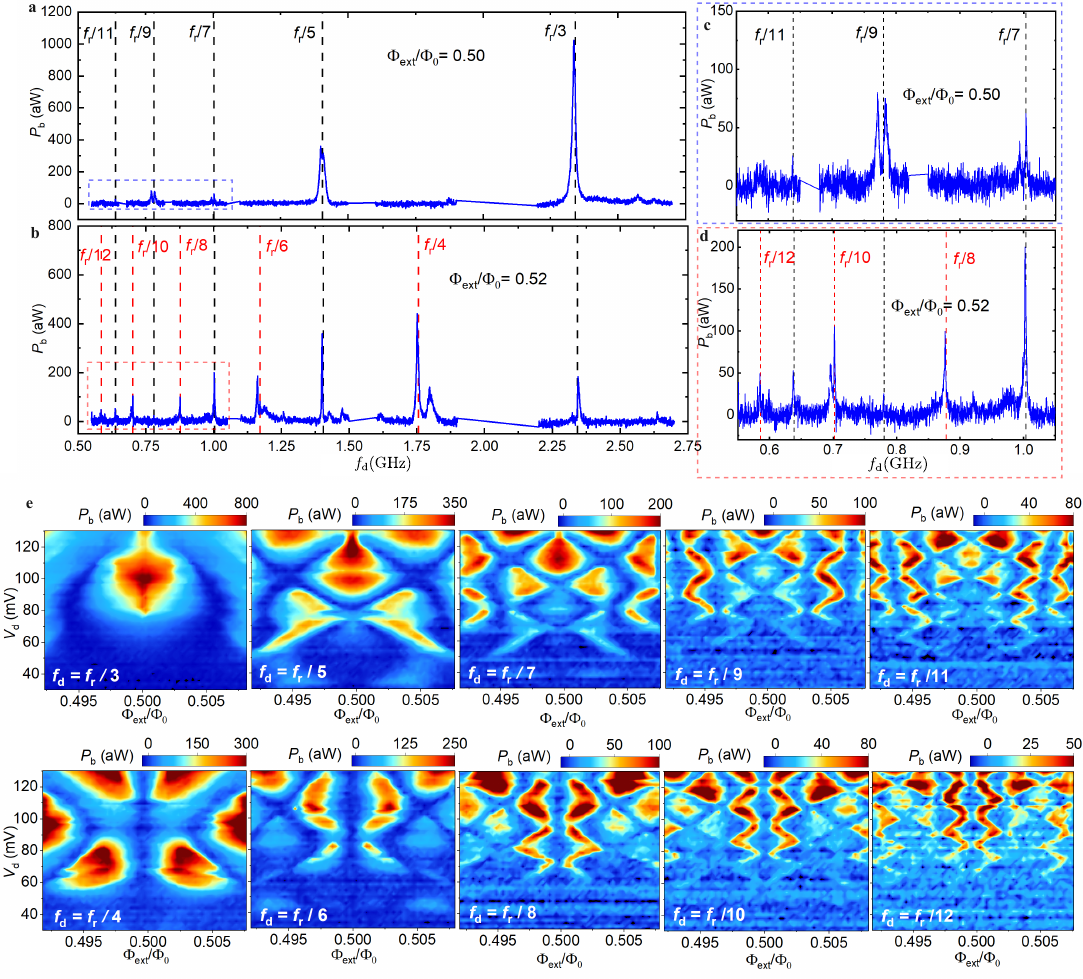}
\caption{\textbf{Heat transfer at fractional frequencies}. \textbf{a}, At the symmetry point, $\Phi_{\text{ext}}/\Phi_0 = 0.5$, heat transfer occurs at driving frequencies $f_{\text{d}} = f_{\text{r}}/l$, where $l$ is an odd integer number due to selection rules. Here $f_{\text{r}}=7.025~\text{GHz}$ is the bare resonator frequency. \textbf{b}, At flux bias $\Phi_{\text{ext}}/\Phi_0 \neq 0.5$, the bias symmetry is broken, and heat transfer occurs at both even and odd integer fractions of $f_{\text{r}}$. The drive amplitude is fixed at $V_{\text{d}}=100~\text{mV}$. \textbf{c-d}, Zoomed-in data from inside the dashed rectangles of \textbf{(a)} and \textbf{(b)} respectively. \textbf{e}, 2D heat maps of measured heat current as a function of drive amplitude $V_{\text{d}}$ and external flux $\Phi_{\text{ext}}$ at driving frequencies $f_{\text{d}} = f_{\text{r}}/l$, top row for $l$ odd and bottom row for $l$ even.} \label{fig:2}  

\end{figure*}

 Coherence effects in heat transport have so far been widely studied theoretically but they were never subject to direct experimental observation. Here, we directly measure such quantum feature in an experimental setup consisting of a driven qubit-resonator system coupled to an engineered thermal bath. The principle of the experiment is that we cyclically modulate the qubit energy, so that the states of the system interfere, and directly measure the heat exchange between the driven system and the heat bath \cite{pekola_2013_calorimetric,Thomas2023}. We observe interference patterns in the measured heat current, indicating the presence of coherence in heat transport. Based on Floquet theory, we quantitatively capture the experiments using the driven quantum Rabi model (QRM) at weak coupling with the heat bath. The transport mechanism can be understood qualitatively by a simpler model of a driven two-level system (TLS) coupled
to a structured heat bath formed of the resonator and the ohmic bath. In this picture, the resonances in the measured heat current are determined by the relative phase accumulated by the qubit states in one drive cycle.

\begin{figure*}

\includegraphics{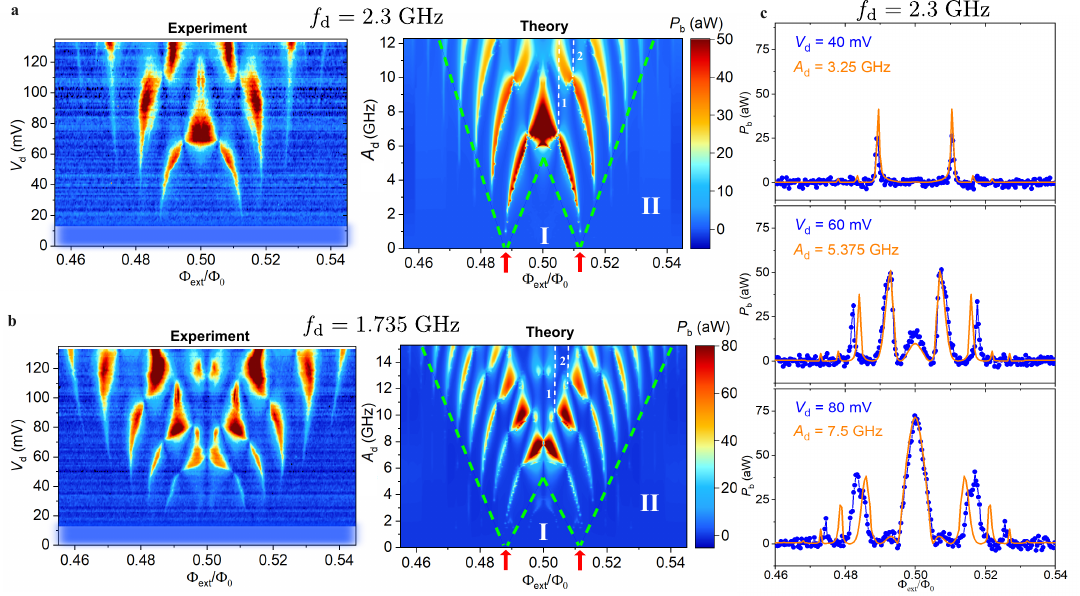}
\caption{\textbf{Quantum interference in heat transport}. Heat current $P_{\text{b}}$ at fixed driving frequency with varying driving amplitude and external flux at \textbf{(a)} $f_{\text{d}}=2.3~\text{GHz}$ and \textbf{(b)} $f_{\text{d}}=1.735~\text{GHz}$. These frequencies roughly correspond to $f_{\text{r}}/3$ and $f_{\text{r}}/4$, respectively. The left panels show the measurement results with varying voltage amplitude $V_{\text{d}}$ and the right panels show the results from simulations using the quantum Rabi model (harmonic oscillator truncated as a $5$-level system) with varying drive amplitude $A_{\text{d}}$. Above the green dashed lines of $W$-shaped "event horizon", the heat current can be detected (above regions I and II). The vertical white dashed lines numbered 1 and 2 in the theory panels mark the conditions $\epsilon_0=f_{\text d}$ and $\epsilon_0=2f_{\text d}$, respectively. \textbf{c}, Comparison between measured (blue dots) and simulated (orange lines) heat current at fixed $f_{\text{d}}=2.3~\text{GHz}$ for three values of the drive amplitude. } \label{fig:3}

\end{figure*}

\section*{The studied system}
We investigate a driven quantum system weakly coupled to a thermal bath as shown in Fig.~\ref{fig:1}(a-b). A flux qubit of frequency $f_{\text{q}}$, and with minimum frequency splitting $\Delta$ at the symmetry point, is cyclically modulated by a continuous sinusoidal driving of frequency $f_{\text{d}}$ and amplitude \(A_{\rm d}\). The qubit is coupled with strength $g$ to a resonator with bare frequency $f_{\text{r}}$ which is in turn weakly coupled, with rate $\gamma$, to a thermal bath of temperature $T_{\text{b}}$. The driving induces transitions in the qubit-resonator levels which eventually relax to the heat bath (transferring heat with average power $P_{\text{b}}$ to the bath via photons). The driven open quantum system is described by the QRM Hamiltonian 
\begin{align}
    &\hat{H}(t)  = -\frac{h}{2}[\Delta \hat{\sigma}_x +  \epsilon(t) \hat{\sigma}_z] + h f_{\text{r}}\hat{a}^{\dagger} \hat{a} + h g (\hat{a}^{\dagger} + \hat{a})\hat{\sigma}_z\label{HamQRM}\\
    &\epsilon(t) = \epsilon_0 + A_{\text{d}} \sin{(2 \pi f_{\text{d}} t)}\;.\label{drive}
\end{align}
The static component of the qubit bias is given by $\epsilon_0=2I_{\text p}(\Phi_{\text{ext}}-\Phi_0/2)/h$, which is the flux detuning of the qubit by an externally applied magnetic flux $\Phi_{\text{ext}}$. Here, $\Phi_0=h/2e$ is the magnetic flux quantum and $I_{{\text p}}$ is the persistent current of the qubit. The time-dependent bias $\epsilon(t)$ modulates cyclically the qubit frequency as \hbox{$f_{\text q}(t)=\sqrt{\Delta^2+\epsilon^2(t)}$}.

Experimentally, the device consists of a quarter-wavelength ($\lambda/4$) coplanar waveguide (CPW) resonator galvanically coupled to the qubit and weakly coupled to the on-chip thermal bath, as shown in the optical microscope of device in Fig.~\ref{fig:1}(c). The qubit is formed of a loop of three Josephson junctions and an inductor \cite{upadhyay_2021_robust} shown in the scanning electron microscope (SEM) image of Fig.~\ref{fig:1}(d). The static qubit frequency can be tuned by a global coil providing the external magnetic flux $\Phi_{\text{ext}}$. The qubit loop is inductively coupled to a nearby on-chip flux bias line connected to a signal generator to inject the voltage drive at amplitude $V_{\textrm{d}}$ and frequency $f_{\textrm{d}}$. This driving induces the ac-magnetic field to the loop and generates the time-dependent term in Eq.~\eqref{drive} with amplitude $A_{\textrm{d}}$ that is proportional to $V_{\textrm{d}}$.

The open-circuit end of the resonator is capacitively coupled to a copper (Cu) island, which is the thermal bath, attached to four normal metal-insulator-superconductor (NIS) junctions functioning as the thermometer to measure the temperature of the bath $T_{\text{b}}$ \cite{Giazotto2006,Satrya2025} as shown in Fig.~\ref{fig:1}(e-f). A fixed current bias $I_{\text{b}}=160~{\text{pA}}$ is applied across the NIS junctions while the voltage drop $V_{\text{th}}$ representing the temperature $T_{\text{b}}$ is monitored. The photonic heat transfer $P_{\text{b}}$ from driven qubit-resonator to the bath (Cu film) can be measured by monitoring the steady-state temperature $T_{\textrm{b}}$ of this mesoscopic bath, see schematic in Fig.~\ref{fig:1}(b). The power can be calculated using the electron-phonon relation \cite{pekola_2021_icolloquiumi,Satrya2025}, since in steady state, the heat flow from electronic island to the phonon bath $P_{\textrm{ep}}$ is equal to the total external heat flow to the island, thus  

\begin{equation}\label{eq:ElectrPhonon}
  P_{\textrm{b}} = \Sigma \Omega (T_{\textrm{b}}^5-T_{0}^5)-P_{\textrm{e}}.
\end{equation}In our setup, the volume and electron phonon constant of the Cu absorber are estimated to be $\Omega = 2.5 ~\textrm{x} ~10^{-20} ~\mathrm{m^{3}}$ and $\Sigma= 2 ~\textrm{x}~ 10^9~\mathrm{WK^{-5}m^{-3}}$\cite{Giazotto2006,PhysRevLett.55.422}, respectively. The temperature of the phonon bath is $T_0\sim50\text{mK}$, and the background environment heating is $P_{\textrm{e}}\sim2~\text{fW}$ which saturates the bath temperature to $T_\text{b}\sim 130~\text{mK}$. Additional parasitic heating is present when we apply the driving field through the flux bias line, where the signal couples directly to the Cu absorber. Details of the layout and measured device can be found in Supplementary Information I. Experimental DC and RF setups are provided in Supplementary Information II.

The resonator is also weakly coupled to a probing feedline for RF spectroscopic characterization of the device. The RF scattering spectroscopy $|S_{21}|$ is performed by injecting a microwave signal to Port-1 and measuring the output from Port-2. Typical one-tone and two-tone characterizations are performed to measure the qubit-resonator spectrum as a function of the applied static external flux $\Phi_{\text{ext}}$ in the absence of time-dependent drives term $A_{\text{d}}\sin{(2 \pi f_{\text{d}}} t)$. The results are shown in Fig.~\ref{fig:1}(g-h) fitted by the Hamiltonian of Eq.~\eqref{HamQRM} (red dashed line). The blue dashed line is the simulation with extra non-linear terms included, see Supplementary Information IV. We obtain the parameters $\Delta = 4.708~\text{GHz}$, $I_\text{p} = 76.61~\text{nA}$, $f_{\text{r}} = 7.025~\text{GHz}$, and $g = 420~\text{MHz}$, which is in the strong coupling regime ($g/f_{\text{r}}\sim6\%$). The estimated static photon-relaxation rate to the bath is $\gamma\approx7~\text{MHz}$ which is small compared to the other frequencies of the system. This justifies the weak coupling assumption between the bath and the quantum system. Detailed spectroscopic characterization results are provided in Supplementary Information IV.

\section*{Heat current measurements}

We measure the voltage drop across a pair of NIS junctions $V_\textrm{th}$ while sweeping the driving parameters $V_\textrm{d}$ and frequency $f_\textrm{d}$. The amplitude $A_\textrm{d}$ is varied from 10 mV to 130 mV from the signal generator while the frequency is swept from 0.5 GHz to 2.75 GHz. The measurements are performed in a dilution refrigerator at around 50 mK temperature. The recorded $V_\textrm{th}$ is converted to bath temperature $T_\textrm{b}$ and power flow $P_\textrm{b}$, according to the procedure described in \cite{Satrya2025}. 

The measured heat current $P_{\text{b}}$ as a function of $f_{\text{d}}$ with fixed $V_{\text{d}}$ = 100 mV is shown in {Fig.~\ref{fig:2}}. Peaks in the transferred heat are observed when the drive frequency matches fractions of the resonator frequency, $f_\textrm{d}=f_\textrm{r}/l$, where $l$ is an integer. The amplitude in this measurement is fixed from the signal generator at room temperature; however the actual amplitude in the sample might vary with the drive frequency due to varying attenuation in the line.

Figure~\ref{fig:2}(a) shows data with the external flux fixed at $\Phi_\textrm{ext}=0.5~\Phi_0$ displaying heat resonance peaks at odd integer values of $l$. This is due to a selection rule emerging at the qubit symmetry point. In contrast, at $\Phi_\textrm{ext}=0.52~\Phi_0$ the resonances occur both at even and odd integer values of $l$, as shown in Fig.~\ref{fig:2}(b). Details of the resonances at lower frequencies are shown in Fig.~\ref{fig:2}(c-d). Note that these heat peaks reach much lower power levels
as compared to the ones at higher frequencies. The 2D heat maps at each fractional frequencies are shown in Fig.~\ref{fig:2}(e), clearly demonstrating that for $l$ even the heat current is zero at the symmetry point $\Phi_\textrm{ext}=0.5~\Phi_0$ regardless of the strength of the drive amplitude. 

The selection rules governing heat transport at $f_\textrm{d}=f_\textrm{r}/l$ and $\Phi_\textrm{ext}=0.5~\Phi_0$, corresponding to $\epsilon_0=0$, can be understood from the effective driven two-level system model where the power is estimated analytically to be $P_{\rm b}\simeq h f_{\text r}G_{\rm eff}(f_{\text r})|Q_{00}^l|^2$, see the Supplementary Information VII. Here the so-called structured spectral density function $G_{\rm eff}$ has a narrow peak centered at the resonator frequency and $Q_{00}^l$ is a Fourier component of the matrix element of the qubit coupling operator in the basis of the Floquet modes. The latter modulates the emitted power, which is zero at the symmetry point when $l$ is an even number. To see this, consider the adiabatic limit, where the Floquet modes are given by the instantaneous eigenstates of the time-dependent Hamiltonian Eq.~\eqref{HamQRM} and their quasienergy separation by the accumulated dynamical phase $\Delta\varepsilon_{\text{ad}}= f_{\text d}\int_0^{1/f_{\text d}} dt\, f_{\text q}(t)$. In this regime, $Q_{00}(t)=\bra {\Phi_0(t)}\hat \sigma_z\ket{\Phi_0(t)}=\epsilon(t)/f_{\text q}(t)$. Expanding $1/f_{\text q}(t)$ as a series, one can see that, for $\epsilon_0=0$, only oscillating terms of frequencies which are odd multiples of $f_{\text d}$ contribute, whereas for $\epsilon_0\neq 0$ also even multiples appear. This selection rule can be also seen from solving the master equation of a simple case of driven qubit coupled to a bath, see Supplementary Information IX. As one can see in Fig.~\ref{fig:2}(a-b), besides the main peaks at the fractional frequencies, several side peaks are present, especially away from the symmetry point. These correspond to the different ways the resonance condition for heat transport, see Methods, is achieved (details are provided in Supplementary Information VI and VII). Similar results for resonant peaks of the excited-state population of a driven qubit at integer fractions of the qubit frequency, along with the suppression of the even peaks for zero static bias, were previously studied in~\cite{Shytov2003,Shevchenko2005,Deppe2008}. 

To see more quantitatively the effects of quantum interference on heat transfer, we show, in Fig.~\ref{fig:3}, the measured and calculated heat current (the simulations are performed using Floquet theory applied to the full quantum Rabi model).
The heat current is shown as a function of the applied bias flux and of the amplitude of the drive at around the resonance condition $f_{\text d}\simeq f_{\text r}/3$, Fig.~\ref{fig:3}(a), and  with the drive frequency close to the even-fraction peak $f_{\text d}\simeq f_{\text r}/4$, Fig.~\ref{fig:3}(b). The latter displays zero heat flux at $\Phi_{\text{ext}}=0.5$, as dictated by the selection rule at the symmetry point. Here in the model, we use the system-bath coupling constant $\alpha=R_{\text b}f_{\text r}C_{\text{b}}^2/2(C+C_{\text{b}})$ with bath resistance $R_{\text b}=12.23~\Omega$, capacitance between resonator and bath $C_{\text b}=19.6$ fF, and equivalent total capacitance of the circuit $C=356$ fF. Figure~\ref{fig:3}(c) shows the quantitative comparison between measured (blue circles) and simulated (orange lines) heat current \emph{vs.} the applied flux at three fixed drive amplitudes for $f_{\text{d}} =2.3~\text{GHz}$. Depending on $\Phi_{\text{ext}}$, the heat flow $P_{\text{b}}$ is modulated from 0 to a maximum value of around 75 aW. We note that, while the features of the measured heat flow are captured by the theory, at large drive amplitude these are shifted in a flux-dependent fashion which could indicate that, for large applied bias $\epsilon_0 + A_{\text d}$, the qubit behavior deviates from the idealized two-level system used in the model, as shown in the two-tone scattering spectroscopy in Fig.~\ref{fig:1}(g).

In the heat maps of Fig.~\ref{fig:3} we identify a $W$-shaped "event horizon" (green dashed line) below which no heat current is detected. This is because below this threshold the amplitude of the drive is not sufficient to achieve resonance between the qubit and the resonator and therefore LZSM transitions between levels separated by the qubit-resonator avoided crossing do not occur ~\cite{Koski2018,Chen2021}. The condition on $A_{\text d}$ as a function of the static bias $\epsilon_0$ which defines this threshold reads $f_{\text r}=\sqrt{\Delta^2+\epsilon_{\rm max}^2}$, where $\epsilon_{\rm max}=\epsilon_0+A_{\text d}$. 
Note that for the two values of bias $\epsilon_0=\pm \sqrt{f_{\text r}^2-\Delta^2}$ that realize the condition for resonant transport in the static case ($A_{\text d}=0$) the heat current is present down to low amplitudes, see the red arrows. In Fig.~\ref{fig:3}(a), above the event horizon, one can single out two regions: the first one is around the symmetry point $\Phi_\textrm{ext}=0.5~\Phi_0$ (or $\epsilon_0=0$) where the heat current increases with increasing amplitude, has a maximum at around $A_{\text d}\sim f_{\text r}$, corresponding to $V_d\sim80$~mV in the experiment, and then is suppressed. At larger applied static flux, arc structures appear corresponding to avoided crossings in the qubit quasienergies: $\Delta\varepsilon=\varepsilon_1-\varepsilon_0\simeq 0,f_{\text d}$ (the case $\Delta\varepsilon= f_{\text d}$ is also considered an avoided crossing due to the periodicity of the quasienergies with period $f_{\text d}$), see also~\cite{Koski2018}. In this situation, the accumulated relative phase in one drive period between different Floquet states is a multiple of $2\pi$ and, again, constructive interference populates both the qubit Floquet states~\cite{Chen2021}. The nodes that separate the arcs form the central region appear at the intersection between the arc of quasidegenerate Floquet states and the vertical lines of $\epsilon_0=m f_{\text d}$. At these points the transition matrix elements vanish, and the qubit dynamics are hindered, preventing heat transfer to the bath, according to the mechanism of coherent destruction of tunneling~\cite{Grossmann1991,Grifoni1998,Stehlik2012}. A similar behavior is observed for $f_{\text d}\simeq f_{\text r}/4$, shown in Fig.~\ref{fig:2}b, except for the vertical line of qubit symmetry point, $\Phi_{\text{ext}}=\Phi_0/2$, where the heat current is suppressed by virtue of the selection rule. An account of this, based on arguments from perturbation theory in the qubit $\Delta$ renormalized by the drive, is provided in Supplementary Information VIII. 

\section*{Conclusion}

We have presented experimentally thermal detection of quantum interference, demonstrating a direct measurement of heat in a quantum thermodynamics experiment. By engineering an open quantum system with a nanobolometer as the heat bath, we performed experiments where thermodynamics and quantum mechanics play key roles. The measurements reveal coherence features, such as interference and selection rules, in heat transfer between the driven qubit-resonator system and the heat bath. The peaks in the heat flux are observed when the drive frequency is an integer fraction of the resonator frequency, resulting from multi-photon processes. Moreover, the peaks at even fractional frequencies are suppressed at the qubit symmetry point. This selection rule is violated by breaking the symmetry upon biasing the qubit with an applied magnetic flux. Our analysis based on Floquet theory captures the main features of the experimental results. In addition, this demonstration is relevant, e.g. to the realization of a cyclic quantum machine such as a quantum Otto refrigerator, where our platform provides the possibility to measure small amounts of heat produced by a driven quantum system, down to few aW resolution. Our results establish a new experimental foundation for studying driven open quantum system where heat exchange is measured directly. 



\section*{Methods}

\subsection*{Theoretical description}

The power to the heat bath in the presence of periodical driving on the qubit can be evaluated using the Floquet-Born-Markov master equation approach~\cite{Kohler1997} that assumes weak system-bath coupling. The driving is treated exactly by projecting the operatorial expressions for the reduced density matrix and the current in the basis of Floquet states~\cite{Grifoni1998} $\ket{\Psi_n(t)}=\exp(-{\rm i}2\pi\varepsilon_n t)\ket{\Phi_n(t)}$, where $\varepsilon_n$ are the quasienergies of the qubit-resonator system and $\ket{\Phi_n(t+1/f_{\text d})}=\ket{\Phi_n(t)}$. For a system with eigenenergies $E_n=h f_n$, the expression for the steady-state heat current $P_{\rm b}$ in the presence of the drive is formally the same as the one in the static case, with the Bohr frequencies $f_{nm}=f_n-f_m$ being replaced by $f_{nm}^l=\varepsilon_n-\varepsilon_m + l f_d$ and the populations and rates expressed in the Floquet basis~\cite{Langemeyer2014,Gasparinetti2014},
\begin{equation}\label{heat_power}
P_{\text{b}}=-\sum_{l,n,m}h f_{nm}^l \Gamma^{(l)}_{nm}p_{m}\;.
\end{equation}
The steady-state populations $p_n$ of the Floquet states are the solution of $\sum_m (\Gamma_{nm} p_m-\Gamma_{mn} p_n)=0$ with $\Gamma_{nm}=\sum_l\Gamma_{nm}^{(l)}$, where the partial rates read
\begin{equation}\label{rates}
\Gamma^{(l)}_{nm}=2\pi |Q_{nm}^{l}|^2 G(f_{nm}^l)n_\text{b}(f_{nm}^l)\;,
\end{equation}
with $n_\text{b}(f)=[e^{hf/k_B T_{\text{b}}}-1]^{-1}$ the Bose-Einstein distribution at the bath temperature $T_{\text{b}}$. Here, $G(f)$ is the bath spectral density function. For an ohmic bath, as the resistor connected to the resonator, $G(f)=\alpha 2 \pi f$ , where $\alpha$ is a dimensionless coupling constant. An estimate for $\alpha$ based on the circuit parameters is given in the Supplementary Information V. Expanding the Floquet modes in Fourier series, $|\Phi_n(t)\rangle=\sum_k e^{{\rm i}k 2\pi f_{\text d} t} |\Phi_n^{(k)}\rangle$, one obtains for the $l$-th Fourier component of the matrix element of the system operator that couples to the heat bath  
$Q_{nm}^{l}=\sum_{k=-\infty}^\infty\bra{\Phi_n^{(k)}}\hat{Q}\ket{\Phi_m^{(k+l)}}$. Detailed derivations are provided in the Supplementary Information V and VI. Equation~\eqref{heat_power} is valid for any periodically-driven system weakly coupled to a heat bath under the assumption that the steady-state coherences in the Floquet basis vanish. The generalization to multiple baths is straightforward.

\indent To gain a qualitative understanding it is useful to apply Eq.~\eqref{heat_power} to an equivalent model where  the qubit is the open system and the remaining part of the circuit forms an effective bath which embeds the resonator, see the detailed description in the Supplementary Information VII~\footnote{
Note that while the weak coupling approach can be justified when applied to the full qubit-resonator system, its reliability is questionable for the two-level system interacting with the structured bath, where the effective coupling can be outside the validity regime of the perturbative approach. Nevertheless this description provides qualitatively sensible results.}. This effective-bath mapping yields a spectral density function $G_{\rm eff}(f)$ with a narrow peak centered close to the resonator frequency, and an effective coupling constant which is determined by the qubit-resonator coupling. The procedure is an exact
mapping to a so-called structured heat bath~\cite{Garg1985,Iles-Smith2016}, and presents the advantage of dealing with only two Floquet states and their corresponding quasienergies, {$\varepsilon_n\in\{\varepsilon_0,\varepsilon_1\}$}, within the ``first Brillouin zone" $[-f_{\text d}/2,f_{\text d}/2]$, the other quasi-energies being replicas separated by integer multiples of the drive frequency~\cite{Grifoni1998}.\\
\indent As for the original problem of the driven qubit-resonator system, the power to the bath is still given by Eqs.~\eqref{heat_power} and~\eqref{rates} upon substituting $G(f)$ with $G_{\rm eff}(f)$ and the Floquet states/quasienergies of the full qubit-resonator system with those of the driven qubit alone. Finally, the resonator-bath coupling operator is substituted by the qubit operator that couples to the resonator, $\hat Q=\hat \sigma_z$, see Eq.~\eqref{HamQRM}. 
The resulting picture is one in which transport occurs resonantly, namely when 
\begin{equation}\label{resonance_condition}
\varepsilon_n -\varepsilon_m + l f_{\text d}\simeq f_{\text r}\;,
\end{equation}
due to the suppression of the rates introduced by $G_{\rm eff}(f)$ away from the resonance condition $f=f_{\text r}$. 

Equation~\eqref{heat_power} implies that a contribution to transport comes from transitions between Floquet states in different Brillouin zones, with quasienergies that differ by multiples of the drive frequency (finite $l$ and $n=m$ or $\varepsilon_n=\varepsilon_m$ with $n\neq m$). For these transitions, the resonance condition~\eqref{resonance_condition} gives the simple expression $f_{\text d}=f_{\text r}/l$, which implies that heat transfer can occur at drive frequencies that are fractions of the resonator frequency $f_{\text r}$. This can be seen in terms of the relative phase between Floquet states accumulated in one drive cycle being multiple of $2\pi$ and resulting in constructive interference~\cite{Chen2021,Ivakhnenko2023,Thomas2023}. Given a Floquet state $\ket{\Psi_n(t)}$, the phase accumulated in one drive period $\mathcal{T}_{\text d}=1/f_{\text d}$ is $\phi_n =2\pi\varepsilon_n \mathcal{T}_{\text d}$. The latter can be split into two contributions: the dynamical and geometric phase. The first is given by the average energy in the Floquet state, namely $\phi_n^{\text D}=(1/\hbar)\int_0^{\mathcal{T}_{\text d}} dt\bra {\Psi_n(t)}\hat H_S(t)\ket{\Psi_n(t)}$  while the geometric contribution $\phi_n^{\text G}=\phi_n-\phi_n^{\text D}$ depends on how the system parameters are varied in a cycle~\cite{Grifoni1998}.

\subsection*{Fabrication}

The fabrication of the device is done in a multistage process on a $675~\mathrm{\mu m}$-thick and highly resistive silicon (Si) substrate, resulting in the device shown in Fig.~\ref{fig:1}(c-f). The fabrication consists of three main steps: (1)  niobium (Nb) structures (resonator, feedline, ground plane, and pads), (2) flux qubit made of three junctions of superconductor–insulator–superconductor (SIS) with aluminium (Al) film, (3) bolometer consisting of absorber and thermometer. A $40~\mathrm{nm}$-thick~Al\textsubscript{}O\textsubscript{x} layer is deposited onto a silicon substrate using atomic layer deposition, followed by a deposition of a $200~\mathrm{nm}$-thick Nb film using DC magnetron sputtering. Positive electron beam resist, AR-P 6200.13, is spin-coated with a speed of 6000~rpm for 60~s, and is soft-baked for 9 minutes at 160$^{\circ}$C, which is then patterned by electron beam lithography (EBL) and etched by reactive ion etching. A shadow mask defined by EBL on a 1~{$\mathrm{\mu m}$}-thick poly(methyl-metacrylate)/copolymer resist bilayer is used to fabricate the flux qubit made of Al film galvanically connected to the resonator at the shorted end \cite{upadhyay_2021_robust}. Before the deposition of the Al film, the Nb surface is cleaned in-situ by Ar ion plasma milling for 60~s. The bolometer structure is fabricated with a three angle deposition technique. To have a clean contact between the Nb and Al films, the Nb surface is cleaned in-situ by Ar ion plasma milling for 45~s, followed by deposition at +40$^{\circ}$ of a 20~nm-thick Al lead. The Al lead is oxidized at 2.5~mbar pressure for 2~minutes. After that, a 3~nm-thick Al buffer layer is deposited at -6.5$^{\circ}$, followed by deposition of 30~nm-thick Cu film at -6.5$^{\circ}$. Finally, a 90~nm-thick Al film is deposited at +20$^{\circ}$ on top of the edge of the Cu film to connect the Cu to the Nb capacitor. After liftoff in hot acetone (52 degrees for ~30 minutes) and cleaning with isopropyl alcohol, the substrate is cut by an automatic dicing-saw machine to $7~\textrm{mm}\times7~\textrm{mm}$ chips. Last, a chip is wire-bonded to copper RF-DC holder for the low-temperature measurements.

\subsection*{Measurement setup}
Measurements are performed in a cryogen-free dilution
refrigerator at temperature of 50 mK with the setup shown and described in Supplementary Information II. To perform RF scattering $S_{21}$ measurements, using a
VNA, a probe microwave tone is supplied to the input feedline
through a 90 dB of attenuation distributed at various
temperature stages of the cryostat. The output probe signal is then
passed through two cryogenic circulators, before being
amplified first by a 40 dB cryogenic amplifier and then by
a 40 dB room-temperature amplifier. For DC thermal measurements, the voltage $V_{\text{th}}$ is measured by a digital multimeter and lock-in amplifier through Thermocoax lines. The device
is mounted in a tight copper holder and covered by an Al shield.

\section*{Data availability}

The data that support the plots within this article are
available from the corresponding author upon reasonable
request.


\section*{Code availability}

The code that support the findings of this study are
available from the corresponding author upon request.


\section*{Acknowledgements}
We thank Milena Grifoni, Joachim Ankerhold, Meng Xu, Mikko Möttönen, Vasilii Vadimov, George Thomas, Ilari Mäkinen and Sergey Shevchenko for fruitful discussions. This work is financially supported by the Research Council of Finland Centre of Excellence programme grant 336810 and grant 349601 (THEPOW). We acknowledge QuantERA II Programme that has received funding from the EU’s H2020 research and innovation programme under the GA No 101017733. B.K. acknowledges funding from the European Union’s Research and Innovation Programme, Horizon Europe, under the Marie Sklodowska-Curie Grant Agreement No. 101150440 (TcQTD). We sincerely acknowledge the facilities and technical support of Otaniemi Research Infrastructure for Micro and Nanotechnologies (OtaNano) to perform this research. We thank VTT Technical Research Center for sputtered Nb films.



\section*{Competing interests}
The authors declare no conflict of interest.

\section*{References}

 \bibliography{MainReferences}

\end{document}


\title{Supplementary Information \\
Heat measurement of quantum interference}

\author{Christoforus Dimas Satrya$^1$}
\email{christoforus.satrya@aalto.fi}

\author{Aleksandr S. Strelnikov$^1$}
\author{Luca Magazz\`u$^{1}$}
 \author{Yu-Cheng~Chang$^{1}$}
 \author{Rishabh Upadhyay$^{1,2}$}

\author{Joonas T. Peltonen$^1$}

\author{Bayan Karimi$^{1,3}$}

\author{Jukka P. Pekola$^1$}
\email{jukka.pekola@aalto.fi}

 \affiliation{$^1$ Pico group, QTF Centre of Excellence, Department of Applied Physics, Aalto University School of Science, P.O. Box 13500, 00076 Aalto, Finland}

\affiliation{$^2$ VTT Technical Research Centre of Finland Ltd, Tietotie 3, 02150 Espoo, Finland}

 \affiliation{$^3$ Pritzker School of Molecular Engineering, University of Chicago, Chicago IL 60637, USA}

\maketitle

\section{Device design and parameters}

\begin{figure*}[ht]
\includegraphics{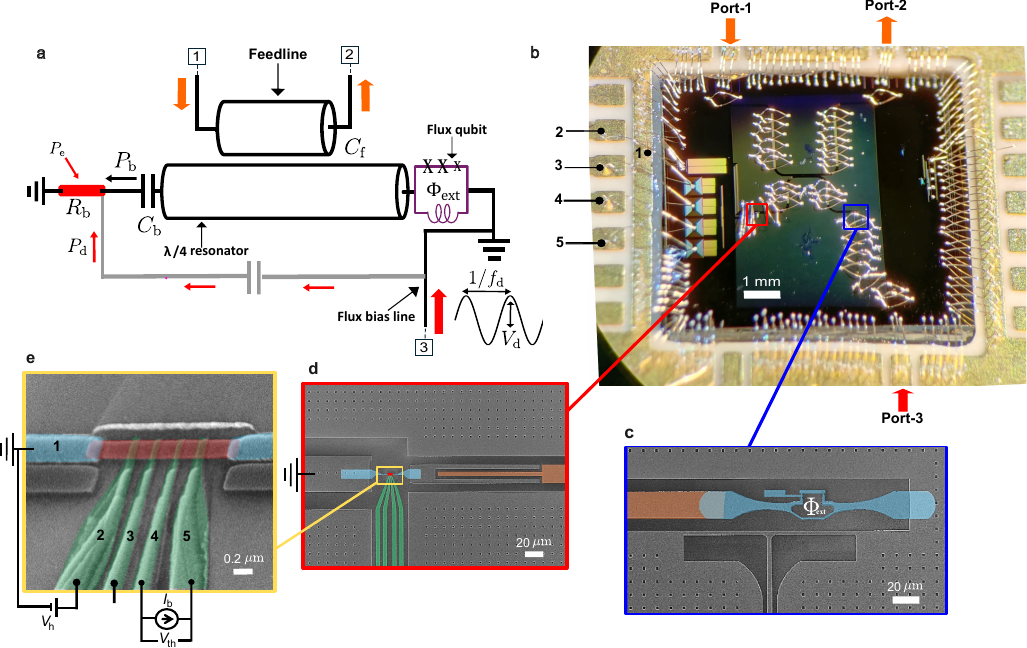}
\caption{\textbf{The schematic and measured device}. \textbf{a}, Circuit schematic of the device consisting of galvanically connected qubit-resonator weakly coupled to probing feedline and bolometer (resistor). \textbf{b}, Optical micrograph of the measured device with the size 7 mm x 7 mm on the sample holder. The open side of the resonator is weakly capacitively coupled to an on-chip thermal bath made of a copper (Cu) film (inside red rectangle). The resonator is weakly coupled also to a feedline for RF spectroscopic measurements, with input line Port-1 and output line Port-2. The flux bias line is connected to Port-3 line. \textbf{c}, SEM image of the flux qubit shunting the resonator. \textbf{d}, SEM image of the open side of resonator (orange) that is capacitively connected to the bolometer (inside yellow rectangle) \textbf{e}, SEM image of the bolometer: the Cu film (red) is connected to four superconducting Al leads (green) with an insulating layer, forming NIS junctions. Line 1 is shunted to the ground on the sample holder's PCB, while lines 2-4 are connected to DC lines Thermocoax extending from the cryostat's mixing chamber stage to a room temperature feedthrough flange. \label{fig:1}}   
\end{figure*}

Here we describe the detailed layout of the device and the fabricated sample. The schematic circuit of the device is shown in Fig.~S\ref{fig:1}(a). A $\lambda/4$ coplanar-waveguide (CPW) resonator is weakly capacitively connected to a probing CPW feedline with a capacitance $C_\text{f}$. The open side of the resonator (left) is weakly capacitively attached to a bolometer with capacitance $C_\text{b}$. The shorted side of the resonator (right) is galvanically shorted by a flux qubit composed of a parallel circuit of Josephson junctions and inductance. The device parameters are listed in Tab.~\ref{table:1}. The equivalent capacitance and inductance are determined by $C=1/(8Z_0f_{\text{r}})$ and $L=2Z_0/(\pi^2f_{\text{r}})$ \cite{Pozar}, where $f_{\text{r}}$ is measured resonance frequency of the resonator. The resistance $R_{\text{b}}$ is obtained from a low-temperature four probe IV measurement. The capacitances $C_{\text{b}}$ and $C_{\text{f}}$ are obtained from COMSOL simulations. The internal loss $R$ in the resonator is calculated by $R=(4 Q_{\text{in}} Z_0) /\pi$ \cite{Pozar}, where $Q_{\text{in}}\sim3000$ is the measured internal quality factor of the resonator in the single photon regime.

\begin{table}[h!]
\centering
\begin{tabular}{||c c||} 
\hline
Parameter & Value\\ [0.5ex] 
\hline\hline
Inductance per unit length, $L_\text{l}$ & $405~\mathrm{nH/m}$ \\ 
 Capacitance per unit length, $C_\text{l}$ & $171~\mathrm{pF/m}$  \\
Length of the resonator, $l$  & $3931~\mathrm{\mu m}$  \\
 Capacitance to feedline, $C_\text{f}$ &  $13.85~\mathrm{fF}$ \\ 
 Capacitance to bolometer, $C_\text{b}$ &  $19.60~\mathrm{fF}$ \\
Bolometer resistance, $R_\text{b}$ &  $12.23~\mathrm{\Omega}$ \\ 
Equivalent capacitance, $C$ & $356~\mathrm{fF}$ \\ 
Equivalent inductance, $L$ & $1.44~\mathrm{nH}$ \\
Internal resistance, $R$ & $1.91\mathrm{x}10^5~\mathrm{\Omega}$ \\
Quality factor to feedline, $Q_\text{f}$  & $1681$  \\
Quality factor to bolometer, $Q_\text{b}$ &  $1716$ \\ 

\hline
\end{tabular}
\caption{The device parameters}
\label{table:1}
\end{table}

Figure~S\ref{fig:1}(b) shows the optical micrograph (OM) of the whole measured device, which has a size of 7 mm x 7 mm. For RF scattering (transmission) measurements, Port-1 (input) and Port-2 (output) are connected to the feedline. Port-3 (input) is connected to the flux bias line to drive the flux qubit. DC line 1 is shunted to the low temperature ground on the holder's PCB. DC lines 2-4 are connected to Thermocoax wiring for DC measurements. The measurement setup for DC and RF, depicted in Fig. S\ref{fig:1}(b), is described in the next section. 

\section{Experimental setups and measurements}

\begin{figure*}[t!]
\includegraphics{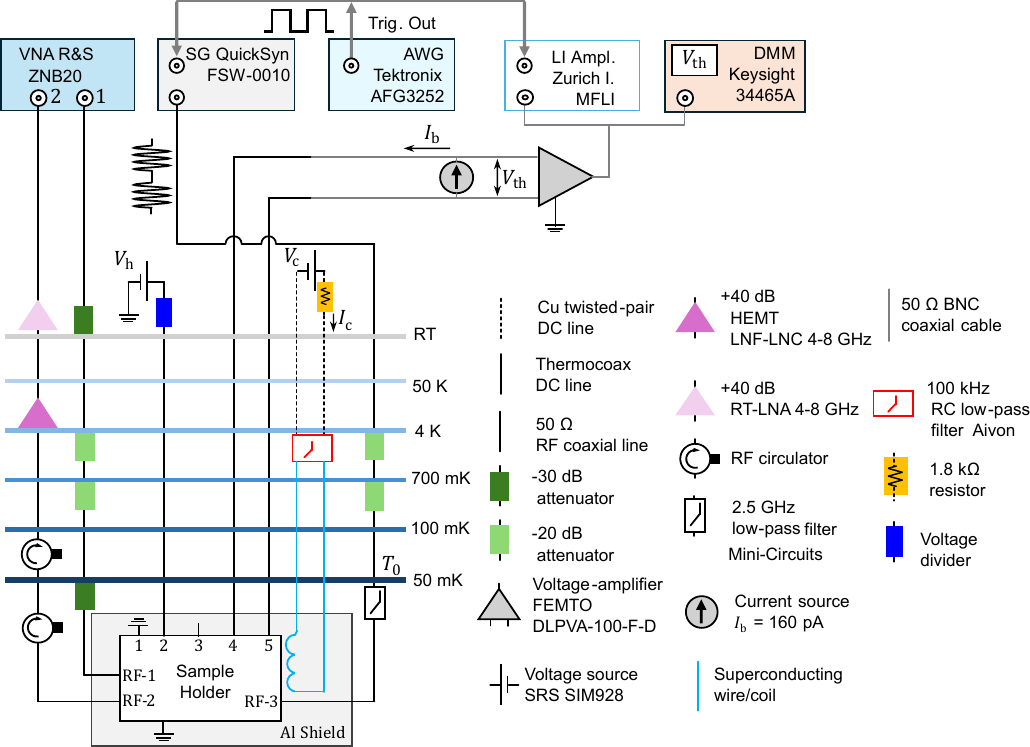}
\caption{\textbf{Measurement setup.}
The RF setup for scattering measurement consists of VNA connected to input line Port-1 and output line Port-2, which are connected to the feedline in the device. The thermal measurement is performed by measuring DC voltage signal across NIS junctions by using multimeter or lock-in amplifier. The device is mounted in a sample holder. The measurements are performed at 50 mK temperature.}  
\label{fig:2}  
\end{figure*}

In this section, we describe the experimental setup used in the experiments. It consists of two parts: the DC setup and the RF setup. The latter is employed for the spectroscopic characterization of the qubit–resonator system, while the DC setup is used for thermal measurements with the on-chip bolometer. The experimental setup is illustrated in Fig.~S\ref{fig:2}. Measurements are conducted in a Bluefors cryogen-free dilution refrigerator with a base temperature at the mixing chamber (MXC) of approximately $50~\mathrm{mK}$. A standard RF setup is employed to measure the scattering parameter $S_{21}$ using a vector network analyzer (VNA). A probe microwave tone is supplied to the feedline to the Port-1 through $90~\mathrm{dB}$ of attenuation, distributed across various temperature stages of the cryostat. Coming out form the Port-2, the probe signal then passes through two cryogenic circulators and is subsequently amplified — first by a $40~\mathrm{dB}$ cryogenic HEMT amplifier, and then by a $40~\mathrm{dB}$ room-temperature amplifier. For two-tone spectroscopy, a second input tone is generated by the VNA’s built-in 2nd signal generator. The DC flux bias is provided by a nearby superconducting coil, which is connected to an isolated voltage source at room temperature. The qubit drive is generated by the QuickSyn signal generator and applied to the on-chip flux bias line (Port-3) through $40~\mathrm{dB}$ of attenuation. The device is housed in a copper holder and enclosed within an aluminum shield. For DC thermal measurements, the SINIS voltage $V_\mathrm{th}$ is measured to probe temperature changes in the Cu film. The voltage is read out either using a digital multimeter (DMM) or a lock-in (LI) amplifier, depending on the signal magnitude. Using the DMM, power resolution as small as $19~\mathrm{aW}$ can be resolved, whereas the LI technique allows sensitivity down to $3~\mathrm{aW}$. The wiring for DC measurements is composed of continuous Thermocoax cables running from the MXC flange to a room temperature feedthrough, thermally anchored to every stage in between.

 As shown in Fig.~S\ref{fig:1}(e), in the measured device, we have four normal metal-insulator-superconductor (NIS) junctions. The normal metal here is Cooper (Cu) film (red) and the superconductor is Aluminum (Al) film (green and blue). The colored green Al films connect to the Cu via an insulator forming NIS junctions with a tunnel resistance $R_{\text{t}}\sim 10~\mathrm{k \Omega}$. The colored blue Al films connect to the Cu via a clean contact; the left side is shunted to the ground whereas the right side is coupled capacitively to the resonator. The labeled junction 2 is used for joule heating the Cu. The labeled junction 3 is unused and floated. The labeled junctions 4 and 5 are used as a pair forming SINIS structure for the thermal readout to measure the Cu electron temperature $T_{\text{b}}$. To do this, a fixed bias current $I_{\text{b}}\sim 160~{\mathrm{pA}}$ is applied accross SINIS and the voltage drop $V_{\text{th}}$ is, first amplified by a room-temperature differential preamplifier, then measured by the DMM Keysight 34465A or digital LI amplifier ZI MFLI. With the calibration function, described in the next section, the measured $V_{\text{th}}$ are converted to temperature $T_{\text{b}}$ and then power $P_{\text{b}}$. Note that, in the heat measurement data $P_{\text{b}}$ presented in the main text, direct parasitic heating from the driving $P_{\text{d}}$ and environment heating $P_{\text{e}}$ are already subtracted, as illustrated in Fig.~S\ref{fig:1}(a).

\section{Thermometer calibration}

\begin{figure*}[ht]
\includegraphics{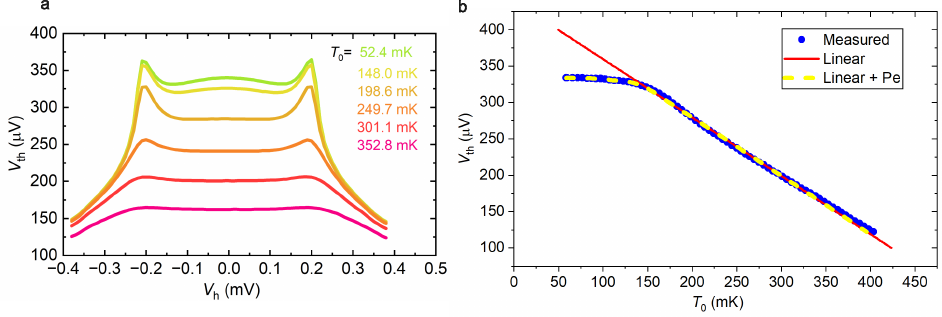}
\caption{\textbf{Thermometry calibration}. \textbf{a}, Thermometer voltage response at different cryostat temperatures. \textbf{b}, Calibration measurement at zero heater voltage ($V_{\text{h}}=0$).\label{figCal}}   
\end{figure*}

In this section we describe the temperature calibration of the bolometer. The calibration is performed to obtain the conversion between voltage $V_{\text{th}}$ and its corresponding electronic temperature $T_\text{b}$. The setup is shown in Fig.~S\ref{fig:1}(e). At various mixing chamber temperatures ($T_{\text{0}}$~=~$50-400~{\text{mK}}$), a heater voltage ($V_{\text{h}}$) is applied across a NIS junction with a tunnel resistance $R_{\text{t}}\sim 9.43~{\mathrm{k \Omega}}$, and $V_{\text{th}}$ is measured across a pair of NIS junctions with applied bias current $I_{\text{b}}=160~\text{pA}$. Due to quasiparticle tunneling, the electronic temperature in the normal metal is elevated or lowered depending on $V_{\text{h}}$ \cite{RevModPhys.78.217,nahum_1994_electronic} as shown in Fig ~S\ref{figCal}(a) for a few different mixing chamber temperatures $T_{\text{0}}$. Figure~S\ref{figCal}(b) shows $V_{\text{th}}$ at $V_{\text{h}}=~0 ~\text{V}$ at different mixing chamber temperatures. We fit the measurement data (blue dots) with a linear function (red line) 
\begin{equation}\label{eq:e4}
  V_{\text{th}} = aT_{\text{0}}+b\;,
\end{equation}where $a=-8.144.10^{-4}$ and $b=4.464.10^{-4}$. The electronic temperature $T_{\text{b}}$ saturates around 130 mK due to parasitic heating load $P_{\text{e}}$ from the environment. The yellow dashed line results from the linear model taking into account non-zero $P_{\text{e}}$. The effective background power $P_{\text{e}}$ varies with the cryostat temperatures; at $T_0$=50 mK, $P_{\text{e}}\simeq 2$~fW. 

Temperature dependence of current and voltage across a NIS junction follows \cite{Feshchenko2015} 

\begin{equation}
    I = \frac{1}{2eR_{\text{t}}} \int_{-\infty}^{\infty}dE n_{\text{S}}(E)[f_{\text{N}}(E-eV) - f_{\text{N}}(E+eV)]\;,
\end{equation}\label{IQPs}where the Fermi distribution is $f_N=1/(1+e^{E/k_{B}T_{b}})$ and the density of states of BCS superconductor is $n_{\text{s}}(E)=|\text{Re}[(E/\Delta + id)/\sqrt{(E/\Delta + id)^2-1}]|$. Here, the superconducting gap of Al film ($\Delta$), Dynes parameter ($d$), and total tunnel resistance of SINIS ($R_{\text{t}}$) are $\Delta \sim 232~\mathrm{\mu eV}$, $d \sim 2.4 ~\textrm{x}~ 10^{-3}$, and $R_{\text{t}} \sim 22.24~{\mathrm{k \Omega}}$, respectively, as obtained from the measurements. 

\newpage

\section{Spectroscopy of the qubit-resonator system}

\begin{figure}[h]
\includegraphics{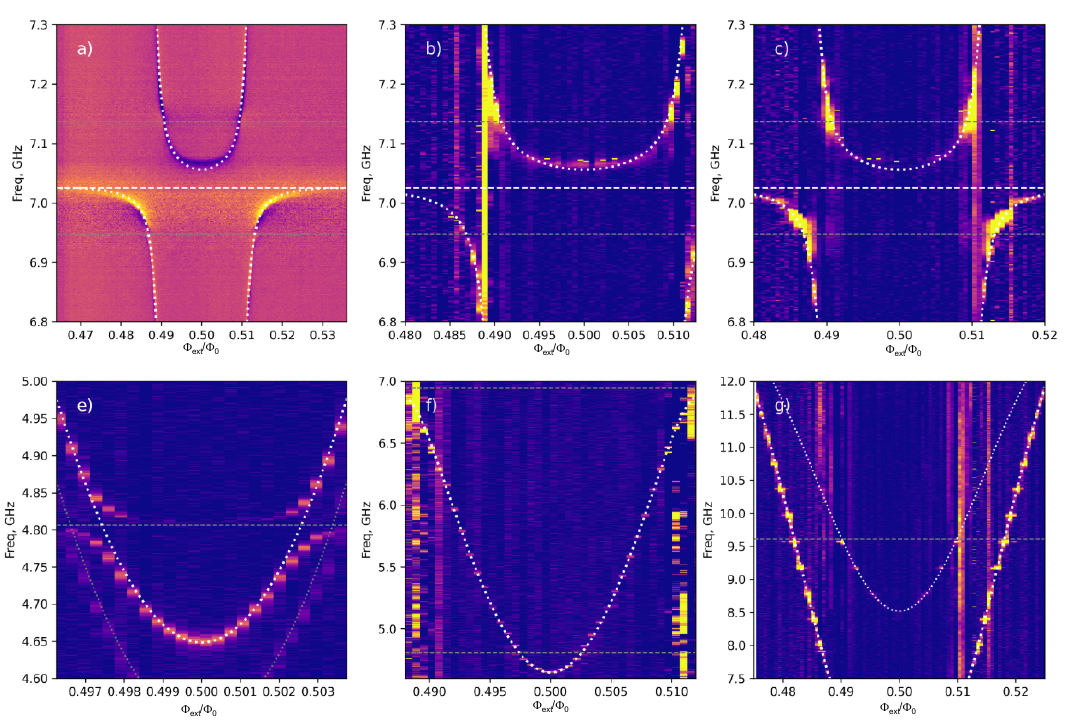}
\centering
\caption{\textbf{Qubit fit result.} Raw datasets: STS a), high-power TTS through flux line c), g), low-power TTS through feedline b), f), high-power TTS through feedline e). White dashed horizontal line - bare resonance frequency, white dotted lines - fit result, gray dashed horizontal lines - parasitic resonances. }
\label{fig:qubit_fit}
\end{figure}

The fit result for the full Hamiltonian is shown in Fig.~S\ref{fig:qubit_fit}. The fit procedure of the qubit-resonator system is the last and most complicated part of the analysis. The system Hamiltonian used consists of two parts: linear, used in the heat current simulations, and nonlinear, caused by system imperfections, such as nonlinear flux dependence the higher qubit levels (here \(f_{(02)}\), Fig.~S\ref{fig:qubit_fit} (g)). Explicitly

\begin{align}
    &H = H_0 + H_{\rm nonlin.}\\
    &H_0/\hbar = \frac12 \Delta \sigma_x + \frac12 \varepsilon \delta f \sigma_z + \omega_r a^{\dagger} a + g (a^{\dagger} + a)\sigma_z\\
    &H_{\rm nonlin.} = \frac12\Delta^{(02)} \sigma_x^{(02)} + \frac12\varepsilon^{(02)} \delta f \sigma_z^{(02)} + g^{(02)} (a^{\dagger} + a)\sigma_z^{(02)} + \\
    &+ \frac12 \varepsilon^{(3)} \delta f^3 \sigma_z + 3 g \frac{\varepsilon^{(3)}}{\varepsilon} \delta f^2(a^{\dagger} + a)\sigma_z + \\
    &+ \frac12 \varepsilon^{(5)} \delta f^5 \sigma_z + 5 g \frac{\varepsilon^{(5)}}{\varepsilon} \delta f^4(a^{\dagger} + a)\sigma_z +\\
    &+ \frac12 \varepsilon_{(02)}^{(3)} \delta f^3 \sigma_z^{(02)} + 3 g^{(02)} \frac{\varepsilon_{(02)}^{(3)}}{\varepsilon_{(02)}} \delta f^2(a^{\dagger} + a)\sigma_z^{(02)}\;. \\
\end{align}

There are also features, not included to fit, such as parasitic resonances of frequencies: 4.806 GHz, 9.612 GHz (4.806 GHz second harmonics), 6.947 GHz, 7.137 GHz; their sidebands and qubit \(f_{12}\) level, that is not visible on spectroscopy.

Sidebands with frequencies \(f_{01} + 7.0252 - 7.137\) and \(f_{01} + 6.947 - 7.0252\) cause qubit line splitting at high power of second tone (Fig.~S\ref{fig:qubit_fit}(e)). The two-photon blue sideband \((f_{01} + 9.612)/2\) nearby half-flux is close to the resonator's frequency, that can cause an additional dispersive shift. Also, the \(f_{12}\) qubit level, having highest frequency around 3.8 GHz and coupling to resonator in the \(\leq\) 100 MHz range can cause a dispersive shift of the resonator, especially nearby half-flux~\cite{Yamamoto_2014}. These factors are the main sources of systematic errors and fit deviation from the observed values nearby the sweet spot.

Coupling to the 4.806 GHz mode is compensated in the row data nearby half flux using the approximate value of dispersive shift \(\chi = g_p^2/(f_{01} - f{p})\) (Fig.~S\ref{fig:qubit_fit}(e)). The data nearby all other parasitic resonances were excluded.

The row data used for fitting consists of four datasets, namely: STS (Fig.~S\ref{fig:qubit_fit}(a)), high-power TTS through flux line (Fig.~S\ref{fig:qubit_fit}(c, g)), low-power TTS through feedline (Fig.~S\ref{fig:qubit_fit}(b, f)), high-power TTS through feedline (Fig.~S\ref{fig:qubit_fit}(e)). It is important to take into account that the zero-flux point for each dataset is slightly shifted, so these shifts are included to the fit function as fit parameters. Such shift breaks data symmetry with respect to the half flux point, so fit of uncompensated data with symmetric fit function cause significant systematic errors.

For all datasets, except the STS, the system levels were extracted with a simple maximum search procedure. STS was fitted with a similar procedure as the bare resonator, but with subtraction of background, that was constructed with data in the \(< 0.46\) flux region, away from bare resonator frequency and in the avoiding-crossing region, nearby bare resonator frequency. The small constant level shift, attributed to amplification creep during measurement (20 h long), was subtracted manually.

\begin{figure}[ht!]
\includegraphics{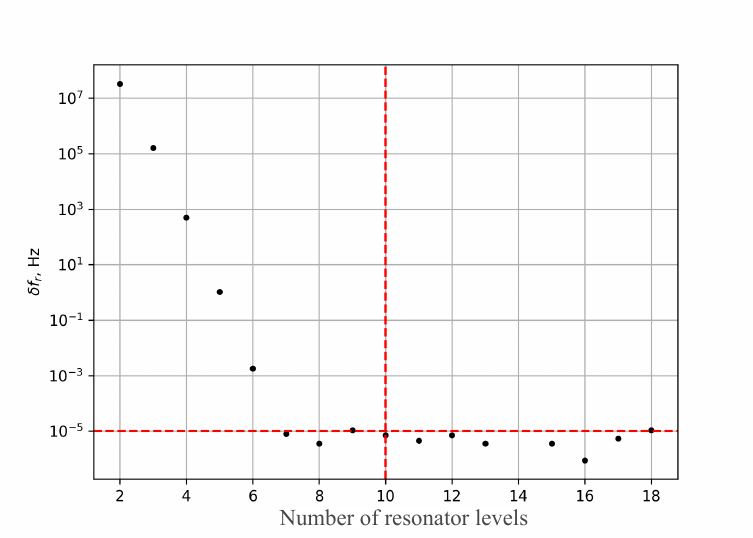}
\centering
\caption{\textbf{Spectrum convergence at half-flux point}}
\label{fig:N1_convergence}
\end{figure}

The number of resonator levels of resonator also plays a role in the fitting procedure, allowing for taking into account the non-RWA terms of \(H_0\). The convergence of the resonator levels number at half flux is shown in Fig.~S\ref{fig:N1_convergence}. In fits, 10 resonator levels were used, enough to achieve a relative tolerance of \(10^{-14}\).

\begin{table}[]
    \centering
    \begin{tabular}{|c|c|c|}
    \hline
         Parameter & Value & Error\\
    \hline
        \(f_\text{r}\) & 7.0252 GHz & 0.0002 GHz\\
    \hline
        \(Q_\text{b}\) & 1000 & 50\\
    \hline
        \(Q_\text{f}\) & 3690 & 50\\
    \hline
    \hline
        \(\varepsilon\) & 478 GHz & 2 GHz\\
    \hline
        \(\varepsilon^{(3)}\) & -33408 GHz & 10063 GHz\\
    \hline
        \(\varepsilon^{(5)}\) & -5.267e7 GHz & 1.333e7 GHz\\
    \hline
        \(\Delta\) & 4.708 GHz & 0.005 GHz\\
    \hline
        \(g\) & 0.420 GHz & 0.008 GHz\\
    \hline
    \hline
        \(\varepsilon^{(02)}\) & 465 GHz & 11 GHz\\
    \hline
        \(\varepsilon_{(02)}^{(3)}\) & -1.6e5 GHz & 1.1e5 GHz\\
    \hline
        \(\Delta^{(02)}\) & 8.490 GHz & 0.011 GHz\\
    \hline
        \(g^{(02)}\) & 0.199 GHz & 0.016 GHz\\
    \hline    
    \end{tabular}
    \caption{Overall fit results}
    \label{tab:cinal resilts}
\end{table}

\newpage

\section{Caldeira-Leggett model and structured bath}

The Caldeira-Leggett model describes an open system bilinearly interacting via the operator $\hat X$ with a heat bath formed of $N$ harmonic oscillators according to the total Hamiltonian~\cite{Caldeira1981,Caldeira1983}
\be{H_Caldeira_Leggett}
\hat H(t)=& \hat H_{\rm S}(t)+
\hat{H}_{B}+\hat{H}_{SB}\\
=&H_{\rm S}(t)+\frac{1}{2}\sum_{j=1}^{N} \left[
\frac{\hat{p}^{2}_{j}}{m_{j}}+m_{j}\omega^{2}_{j}\left ( \hat{x}_{j}
-\frac{c_{j}} {m_{j}\omega^{2}_{j} } \hat{X} \right )^{2}\right].
\ee
The spectral density function $J(\omega)$ fully characterizes the bath and its couplings to the open system. It is defined as
\begin{equation}\label{J}
J(\omega)=\frac{\pi}{2}\sum_{j=1}^{N}\frac{c_{j}^{2}}{m_{j}\omega_{j}}\delta(\omega-\omega_{j})\;.
\end{equation}
Let us introduce the coupling constants with dimension of an angular frequency 
\[
\lambda_j=\frac{x_0}{\sqrt{2\hbar m_j \omega_j}}c_j\;,
\]
where $x_0$ is a reference length scale.
It is customary to define the modified spectral density function $G(\omega)$, with dimensions of frequency, as~\cite{Weiss2012} 
\be{G}
G(\omega):=\sum_{j=1}^{N}\lambda_j^2\delta(\omega-\omega_{j})=\frac{x_0^2}{\pi\hbar}J(\omega)\;.
\ee
In the continuum limit, the spectral density function of a so-called ohmic bath reads $J(\omega)=M\gamma\omega$, where $\gamma$ is the memoryless friction kernel that measures the overall system-bath coupling. This gives $G(\omega)=\alpha\omega$, where $\alpha=M\gamma x_0^2/\pi\hbar$ is the dimensionless system-bath coupling strength.
If the open system coupled to the bath is a harmonic oscillator of frequency $\omega_{\text r}$, we have $x_0=\sqrt{\hbar/2M\omega_{\text r}}$.\\
\indent Using $\hat{q}_j=\sqrt{\hbar/2m_j\omega_j}(\hat b_j+\hat b_j^\dag)$ and defining the dimensionless system position operator $\hat{Q}$ via $\hat{X}=x_0\hat{Q}$, we can write the Hamiltonian~\eqref{H_Caldeira_Leggett} in the standard form
\be{HCL}
\hat H(t)=& \hat H_{\rm S}(t)+\sum_{j=1}^{N}\hbar\omega_j \hat b_j^\dag \hat b_j -\hat{Q}\sum_{j=1}^{N} x_0c_j\sqrt{\frac{\hbar}{2m_j\omega_j}}(\hat b_j+\hat b_j^\dag)+\hat A^2\\
=&\tilde{H}_{\rm S}(t)+\sum_{j=1}^{N}\hbar\omega_j \hat b_j^\dag \hat b_j -\hat{Q}\sum_{j=1}^{N} \hbar\lambda_j(\hat b_j+\hat b_j^\dag)\\
=&\tilde{H}_{\rm S}(t)+\hat H_{\rm B} +\hat H_{\rm SB}\;.
\ee
Here, we have defined 
\be{Hterms}
\tilde{H}_{\rm S}(t)=&\;\hat H_{\rm S}(t)+\hat A^2\\
\hat H_{\rm B}=&\;\sum_{j=1}^{N}\hbar\omega_j \hat b_j^\dag \hat b_j\\
\hat H_{\rm SB}=&\;-\hat{Q}\otimes\hat{B}\;,\qquad{\rm with}\qquad \hat{B}:=\sum_{j=1}^{N} \hbar\lambda_j(\hat b_j+\hat b_j^\dag)\;,
\ee
where $\tilde{H}_{\rm S}(t)$ is renormalized by absorbing the term $\hat A^2$ which is proportional to the square of the system operator coupled to the bath. It given by, see Eqs.~\eqref{H_Caldeira_Leggett}-\eqref{G},
\begin{equation}\label{A2}
\hat A^2=\sum_{j=1}^{N}\frac{c_{j}^{2}}{2 m_{j}\omega_{j}^2}\hat{X}^2=\frac{1}{\pi}\int_{0}^{\infty}d\omega\frac{J(\omega)}{\omega}\hat{Q}^2x_0^2 = \hbar\int_{0}^{\infty}d\omega\frac{G(\omega)}{\omega}\hat{Q}^2\equiv\mu \hat Q^2 \;.
\end{equation}
For a ohmic spectral density function with Drude cutoff at frequency $\omega_D$, i.e.  \hbox{$G(\omega)=\alpha\omega/[1+(\omega/\omega_D)^2]$}, we have $\mu=(\pi/2)\alpha\hbar\omega_D $.  
Note that if the system operator coupling to the bath is a qubit's Pauli operator, e.g. $\hat{Q}=\sigma_z$, then $\hat{Q}^2$ is proportional to the identity and the term $\hat A^2$ merely constitutes a constant shift in the Hamiltonian.\\

\subsection{Mapping to a structured heat bath}
\label{mapping}

\indent Consider a system coupled to the bath via a harmonic oscillator, with bath spectral density function $G(\omega)=\alpha\omega$, where $\alpha:= \gamma/2\pi\omega_{\text r}$. The dissipative oscillator coupled to the system can be mapped into a structured bath, peaked at the oscillator frequency $\omega_{\text r}$, with effective spectral density function~\cite{Garg1985,Iles-Smith2014,Iles-Smith2016} 
\be{}
G_{\rm eff}(\omega)=\frac{4\alpha g^2\omega_{\text r}^2\omega}{(\omega_{\text r}^2 - \omega^2)^2 + (\gamma\omega)^2}\;.
\ee
Note that in the limit $\alpha\rightarrow 0$, namely disconnecting the dissipative bath, we get $G_{\rm eff}(\omega)\rightarrow \lambda^2\delta(\omega-\omega_{\text r})$, i.e. the system is coupled with coupling strength $\lambda=g$ to a single oscillator of frequency $\omega_{\text r}$.\\

\subsection{Relations with the parameters in a circuit-QED experiment}\label{effective_circuit}

In the experiment, the flux qubit is coupled to a waveguide resonator, modeled as a parallel RLC circuit with large resistance $R$ (in parallel). This circuit is, in turn, capacitively coupled, via $C_{\text b}$, to the resistance $R_{\text b}\ll R$. The parameters are $C=356$ fF, $L=1.44$ nH, $R=1.91\times 10^5~\Omega$, $C_{\text b}=19.6$ fF,  $R_{\text b}=12.23~\Omega$.
\begin{figure}[ht!]
\begin{center}
\includegraphics[width=0.45\textwidth,angle=0]{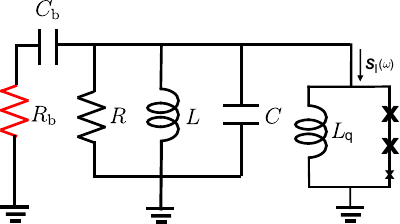}
\caption{Simplified circuit scheme of the resistor-resonator-qubit setup.}
\label{Qubit-resonator-resistor}
\end{center}
\end{figure}

%
The capacitive coupling between the bath and the harmonic oscillator is of the momentum-momentum type. According~\cite{Cattaneo2021} (see Eq. (63) therein) for a ohmic resistor
$$H_{SB}=(\hat a^\dag - \hat a)\frac{\lambda C_{\text b}}{C_{\text b} + C}\sum_{j}\hbar\tilde\lambda_j(\hat b_j^\dag - \hat b_j)\;,$$  
where $\lambda=\sqrt{\hbar\omega_{\text r} (C+C_{\text b})/2}$ and $\tilde\lambda_j=\sqrt{R_{\text b}\omega_j\Delta\omega/\pi\hbar}$. Here, $\omega_j=j\Delta\omega$ are the frequencies
of the bath modes which form a continuum in the limit $\Delta\omega\rightarrow 0$. Using the overall coefficients $\lambda_j=\lambda\tilde\lambda_j$ in the definition~\eqref{G} of spectral density function and assuming for the continuum limit the ohmic form $G(\omega)=\alpha\omega\Theta(\omega)$ with a high-frequency cutoff $\Theta(\omega)$ we obtain for the coupling constant
\[
\alpha=\frac{R_{\text b}\omega_{\text r}}{2}\frac{C_{\text b}^2}{C+C_{\text b}}=\frac{1}{2Q}\frac{C_{\text b}^2}{C(C+C_{\text b})}\simeq 3\times 10^{-4}\;.
\]
Note that the value $\alpha=3.75\times 10^{-4}$ used in the simulations of the full driven quantum Rabi model is close to this value.  

According to~\cite{Goeppl2008} the parallel LC circuit representing the resonator capacitively coupled to the resistor can be represented by a lumped element where the LC circuit, the capacitance and the resistor are in parallel. The resulting impedance is
\be{}
Z(\omega)=&\left(\frac{1}{\ii\omega L}+\ii\omega C+\frac{1}{R}+\ii\omega C^*+\frac{1}{R^*}\right)^{-1}=\frac{\tilde R}{1+\ii \tilde Q \left(\frac{\omega}{\tilde\omega_{\text r}}-\frac{\tilde \omega_{\text r}}{\omega}\right)}\;,
\ee
where $\omega_{\text r}=1/\sqrt{LC}$ and
\be{TLSparameters}
\tilde Q&=\frac{\tilde R}{2\pi}\sqrt{\frac{C+C^*}{L}}\;,\qquad \tilde R =\frac{RR^*}{R+R^*}\;,\qquad \tilde \omega_{\text r}=\frac{1}{\sqrt{L(C+C^*)}}<\omega_{\text r}\;,\\
C^* &=\frac{C_{\text b}}{1+\omega_{\text r}^2 C_{\text b}^2 R_{\text b}^2}\;,\qquad R^*=\frac{1+\omega_{\text r}^2 C_{\text b}^2 R_{\text b}^2}{\omega_{\text r}^2 C_{\text b}^2 R_{\text b}}\;.
\ee
Correspondingly, the spectrum of the current fluctuations across this effective dissipative circuit is given by~\cite{Vool2017}
\be{}
S_{I}(\omega)=2\pi \hbar^2 G_{\rm eff}(\omega)(n_{\text b}(\omega)+1) \propto S_{\phi}(\omega)=\frac{1}{\omega^2}S_{V}(\omega)= \frac{\hbar}{\omega}\frac{2{\rm Re}[Z(\omega)]}{1-e^{-\beta\hbar\omega}}\;.
\ee
With this choice, the spectral density function of the effective bath reads
\be{Geff}
G_{\rm eff}(\omega)=\frac{\tilde\alpha/\omega}{1+\tilde Q^2 \left(\frac{\omega}{\tilde\omega_{\text r}}-\frac{\tilde \omega_{\text r}}{\omega}\right)^2}\;,
\ee
where $\tilde\alpha$, with dimensions of a frequency squared, depends on $\tilde R$ and on the coupling between the qubit and the remaining part of the circuit. 
With these experimental parameters, we have $\tilde Q\simeq 180$ and $\tilde R \simeq 7\times 10^4~\Omega$. Similarly as for the full quantum Rabi model simulations, in the two-level system spectral density function we use a lower effective quality factor to account for the effect of the feed line, which is not present in our theoretical model, see Fig.~S\ref{simulationf3}.

\section{Floquet theory of heat transport in a driven quantum system}

In order to be consistent with the notation used the cited literature on the Floquet approach, in what follows we use angular frequencies in place of frequencies. This is in contrast with the main text where instead, the system parameters and the Floquet quasienergies are expressed in terms of frequencies and with units of GHz.\\
\indent Consider a periodically driven system, with Floquet quasienergies $\hbar\varepsilon_n$ and corresponding Floquet states $\ket{\Psi_n(t)}=\exp(-\ii\varepsilon_n t)\ket{\Phi_n(t)}$, where the Floquet mode $\ket{\Phi_n(t)}=\sum_{k=-\infty}^\infty e^{\ii k \omega_{\text d} t}\ket{\Phi_n^{(k)}}$ is a periodic function of $t$ with the period of the drive $2\pi/\omega_{\text d}$. The system is coupled to a bosonic heat bath with spectral density function $G(\omega)$ and temperature $T_{\rm b}$ via the system coupling operator $\hat Q$.
As shown in Appendix~\ref{Floquet-Redfield} and according to \cite{Langemeyer2014,Gasparinetti2014}, in the weak-coupling, secular master equation approach, the heat transferred to the heat bath is given by  
\be{J}
P_{\rm b}=-\sum_{l,n,m}\hbar\omega_{nm}^l \Gamma^{(l)}_{nm}p_{m}\;.
\ee
Here, $p_m=\rho_{mm}^\infty$ is the occupation probability of the Floquet state $\ket{\Psi_m(t)}$ at the steady state and, assuming $G(-\omega)=-G(\omega)$~\footnote{This extension to negative frequencies holds for Ohmic baths and the corresponding effective, structured spectral density functions.},
\be{definitions}
\omega_{nm}^l&=\varepsilon_n - \varepsilon_m + l\omega_{\text d}\\
\Gamma^{(l)}_{nm}&=2\pi|Q_{nm}^{l}|^2 G(\omega_{nm}^l)n_{\text b}(\omega_{nm}^l)=\frac{1}{\hbar^2}|Q_{nm}^l|^2 S_B(-\omega_{nm}^{l})\\
Q_{nm}^{l}&=\sum_{k=-\infty}^\infty\bra{\Phi_n^{(k)}}\hat{Q}\ket{\Phi_m^{(k+l)}}\;,
\ee
where $l\in(-\infty,\infty)$ and where $n_{\text b}(\omega)=[\exp(\hbar\omega/k_{\rm B}T_{\rm b})-1]^{-1}$ is the Bose-Einstein distribution with the property $n_{\text b}(-x)=-[n_{\text b}(x)+1]$ and $$S_{B}(\omega)= \int_{-\infty}^{\infty}dt\langle \hat{B}(t)\hat{B}(0)\rangle 
e^{\ii\omega t}\\
=2{\rm Re}\int_{0}^{\infty}dt\langle \hat{B}(t)\hat{B}(0)\rangle 
e^{\ii\omega t}$$
is the noise power spectrum of the bath. Details are provided in Appendix~\ref{Floquet-Redfield}.\\

The Floquet-state populations are solutions of the secular master equation 
\be{MEpopulations}
0=\sum_m (\Gamma_{nm} p_m-\Gamma_{mn} p_n)\;,\qquad \Gamma_{nm}=\sum_l\Gamma_{nm}^{(l)}\;,
\ee
see~\cite{Grifoni1998,Kohler2024}. 
These results can be obtained by using the full counting statistic approach for the calculation of the heat current or by projecting in the basis of Floquet states the operatorial expressions from the Nakajima-Zwanzing approach to lowest order.
\comment{
\RED{ALTERNATIVE
\be{}
P_{\rm b}=-\sum_{l,n,m}\hbar\omega_{nm}^l \Gamma^{(l)}_{mn}p_{m}\;.
\ee
Here, $p_m=\rho_{mm}^\infty$ is the occupation probability of the Floquet state $\ket{\Phi_m(t)}=\sum_{k=-\infty}^\infty e^{\ii k \omega_{\text d} t}\ket{\Phi_m^{k}}$ at the steady state and
\be{}
\omega_{nm}^l&=\varepsilon_n - \varepsilon_m + l\omega_{\text d}\\
\Gamma^{(l)}_{mn}&=\frac{1}{\hbar^2}|Q_{mn}^l|^2 S_B(\omega_{mn}^{l})\\
Q_{nm}^{l}&=\sum_{k=-\infty}^\infty\bra{\Phi_n^{(k)}}\hat{Q}\ket{\Phi_m^{(k+l)}}\;,
\ee
where $l\in(-\infty,\infty)$. Here
$$S_{B}(\omega)= \int_{-\infty}^{\infty}dt\langle \hat{B}(t)\hat{B}(0)\rangle 
e^{\ii\omega t}\\
=2\pi\hbar^2 G(\omega)[n_{\text b}(\omega)+1]$$
is the noise power spectrum of the bath with $n_{\text b}(\omega)=[\exp(\beta \hbar\omega)-1]^{-1}$ the Bose-Einstein distribution and we are assuming the extension to negative frequencies with the property $G(-\omega)=-G(\omega)$.  Details are provided in Appendix~\ref{Floquet-Redfield}.\\
The Floquet-state populations are solutions of the secular master equation 
\be{}
0=\sum_m (\Gamma_{mn} p_m-\Gamma_{nm} p_n)\;,\qquad \Gamma_{nm}=\sum_l\Gamma_{nm}^{(l)}\;,
\ee
see}~\cite{Grifoni1998,Kohler2024}. \RED{These results can be obtained by using the full counting statistic approach for the calculation of the heat current or by projecting in the basis of Floquet states the operatorial expressions from the Nakajima-Zwanzing approach to lowest order.}
}
%
%
%

\section{Driven Rabi Hamiltonian and effective model}
\label{}
The quantum Rabi model (QRM) describes a two-level system (TLS), namely a qubit, with tunneling splitting $\Delta$ and bias $\epsilon$, coupled to a harmonic oscillator of frequency $\omega_{\text r}$.
In the presence of a periodic driving on the qubit and in the so-called qubit basis $\{\ket{\pm}\}$, the basis of states with clockwise/counterclockwise persistent current, the Hamiltonian of the driven QRM has the form
\be{HdrivenQRM}
\hat{H}_{\rm QRM}(t)=-\frac{\hbar}{2}\left[\Delta\hat \sigma_x + (\epsilon_0 + A_{\text d}\cos(\omega_{\text d} t))\hat \sigma_z \right] +\hbar\omega_{\text r} \hat a^\dagger \hat a+\hbar g\hat \sigma_z \left(\hat a^\dagger + \hat a \right)\;,
\ee
where $\hat \sigma_z=\ket{+}\bra{+}-\ket{-}\bra{-}$ and $\hat \sigma_x=\ket{+}\bra{-}+\ket{-}\bra{+}$ and where $\hat a^\dag$ ($\hat a$) creates (annihilates) an excitation in the harmonic oscillator.  The angular frequency $g$ quantifies the coupling strength.\\
\indent If the qubit-oscillator system is coupled to a bosonic heat bath via the oscillator, the total Hamiltonian is given by Eq.~\eqref{HCL} with $\hat{H}_S(t)\equiv \hat{H}_{\rm QRM}(t)$ and the system bath coupling operator is $\hat{Q}=\hat a + \hat a^\dag$. The total Hamiltonian is then
\be{}
\hat H(t)= \hat{H}_{\rm QRM}(t)+\hat{A}^2 -\hat{Q}\sum_{j=1}^{N} \hbar\lambda_j(\hat b_j+\hat b_j^\dag)+\sum_{j=1}^{N}\hbar\omega_j \hat b_j^\dag \hat b_j\;.
\ee
\indent In the effective model described in Sections~\ref{mapping} and~\ref{effective_circuit} the qubit is coupled to the structured heat bath via $\hat Q=\sigma_z$ and 
\be{}
\hat{H}_{\rm eff}(t)=-\frac{\hbar}{2}\left[\Delta\hat \sigma_x + (\epsilon_0 + A_{\text d}\cos(\omega_{\text d} t))\hat \sigma_z \right]-\hat{Q} \sum_{j=0}^N \hbar\lambda_j(\hat b_j+\hat b_j^\dag)+\sum_{j=0}^N\hbar\omega_j \hat b_j^\dag \hat b_j \;,
\ee
Notice that while the mapping is exact, an approximate approach (e.g. weak coupling master equation) that is justified in the dissipative QRM might not be so when applied to the TLS interacting with a structured bath. An obvious example is when the qubit-oscillator coupling is strong. Indeed, reaction coordinate mapping techniques were introduced to deal with strongly-coupled structured baths by including part of the latter in the open system allowing for a weak coupling treatment~\cite{Woods2014,Iles-Smith2014}.

\comment{
\subsubsection{RWA and adiabatic driving}

For $\omega_{\text d}\ll\Delta$, we assume that the energy levels and eigenstates follow adiabatically the driving. The time-dependent Rabi Hamiltonian reads in the qubit energy basis
\be{HRabi_energy}
H_{\rm Rabi}(t)=& -\frac{\hbar}{2}\omega_{\text q}(t)\hat{\sigma}_z +\hbar \omega_{\text r} \hat a^\dagger \hat a + \hbar\left[g_z(t)\hat{\sigma}_z -g_x(t)\hat{\sigma}_x  \right] \left(\hat a^\dagger + \hat a \right)\;
\ee
with $\omega_{\text q}(t):=\sqrt{\Delta^2+\epsilon^2(t)}$ and the interaction terms with longitudinal and transverse couplings 
$g_z= g \epsilon(t)/\omega_{\text q}(t)$ and $g_x= g \Delta/\omega_{\text q}(t)$.
Within the RWA, the time-dependent energy levels are
\begin{equation}
\begin{aligned}
\label{JCMspectrum}
\omega_0=&-\omega_{\text q}(t)/2\;,\\
\omega_{2n-1}=& \left(n-\frac{1}{2} \right)\omega_{\text r} - \frac{1}{2}\sqrt{\delta^2(t) + n 4 g_x^2(t)}\;, \\
\omega_{2n}=& \left(n-\frac{1}{2} \right)\omega_{\text r} + \frac{1}{2}\sqrt{\delta^2(t) + n 4g_x^2(t)}\;,
\end{aligned}
\end{equation}
($n\geq 1$), where the detuning is defined as $\delta(t):=\omega_{\text q}(t)-\omega_{\text r}$. Note that $g_z(t)$ doesn't enter the expressions for the eigenrequencies. As a result they depend on the driving via the time dependent qubit frequency $\omega_{\text q}(t)$.

Performing the average over one period driving we consider the effective frequency, which is the difference between the qubit quasienergies in the adiabatic regime~\cite{Koski2018}
\be{}
\tilde\omega_{\text q}=\frac{\omega_{\text d}}{2\pi}\int_0^{2\pi/\omega_{\text d}} d t \sqrt{\Delta^2+[\epsilon_0 + A_{\text d}\cos(\omega_{\text d} t)]^2} 
\ee
and correspondingly $\tilde\delta=\tilde\omega_{\text q}-\omega_{\text r}$ and $\tilde g_x= g \Delta/\tilde\omega_{\text q}$

The corresponding eigenstates, in the energy basis of the uncoupled system, are
\begin{equation}
\begin{aligned}
\label{JCMeigenstates}
|0\rangle &= |{\rm g},0\rangle\;,\\
|2n-1\rangle &= \tilde {\rm u}_{n}^-|{\rm e},n-1\rangle + \tilde {\rm v}_{n}^-|{\rm g},n\rangle\;,\\
|2n\rangle &= \tilde {\rm u}_{n}^+|{\rm e},n-1\rangle + \tilde {\rm v}_{n}^+|{\rm g},n\rangle\;,
\end{aligned}    
\end{equation}
with coefficients
\begin{equation}
\begin{aligned}\label{}
\tilde {\rm u}_n^\pm &:=\;\frac{\tilde \delta \pm \sqrt{\tilde \delta ^2 + n4g_x^2}}{\sqrt{\left(\tilde \delta \pm \sqrt{\tilde \delta^2+n4\tilde g_x^2}\right)^2+n4\tilde g_x^2}}\;,\qquad{\rm and}\qquad
\tilde {\rm v}_n^\pm :=\;\frac{-\sqrt{n}2\tilde g_x}{\sqrt{\left(\tilde \delta \pm \sqrt{\tilde \delta^2+n4\tilde g_x^2}\right)^2+n4\tilde g_x^2}}\,.
\end{aligned}
\end{equation}

Within the RWA, the relevant matrix elements of the system's coupling operators ($\hat{Q}_L=\hat{a}+\hat{a}^\dag$ and $\hat{Q}_R=\sigma_z\epsilon/\omega_{\text q}-\sigma_x \Delta/\omega_{\text q}$, in the qubit energy basis) read
\be{}
Q_{L,01}&={\rm v}_1^-\;,\qquad Q_{R,01}={\rm u}_1^-\frac{\Delta}{\omega_{\text q}}\\
Q_{L,02}&={\rm v}_1^+\;,\qquad Q_{R,02}={\rm u}_1^+\frac{\Delta}{\omega_{\text q}}\\
\ee
}

\subsection{TLS coupled to a bath with effective spectral density function: origin of the peaks of the power to the bath}

\begin{figure}[ht!]
\begin{center}
\includegraphics[width=0.752\textwidth,angle=0]{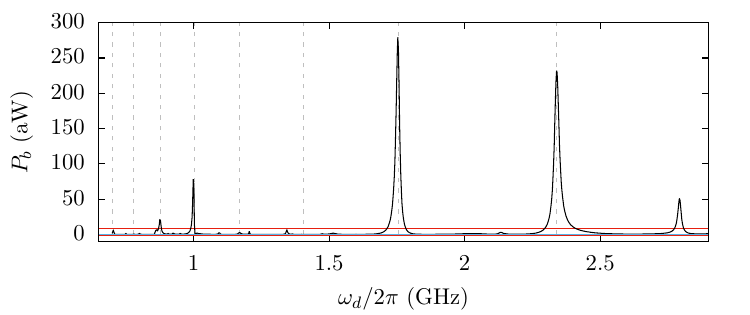}
\includegraphics[width=0.735\textwidth,angle=0]{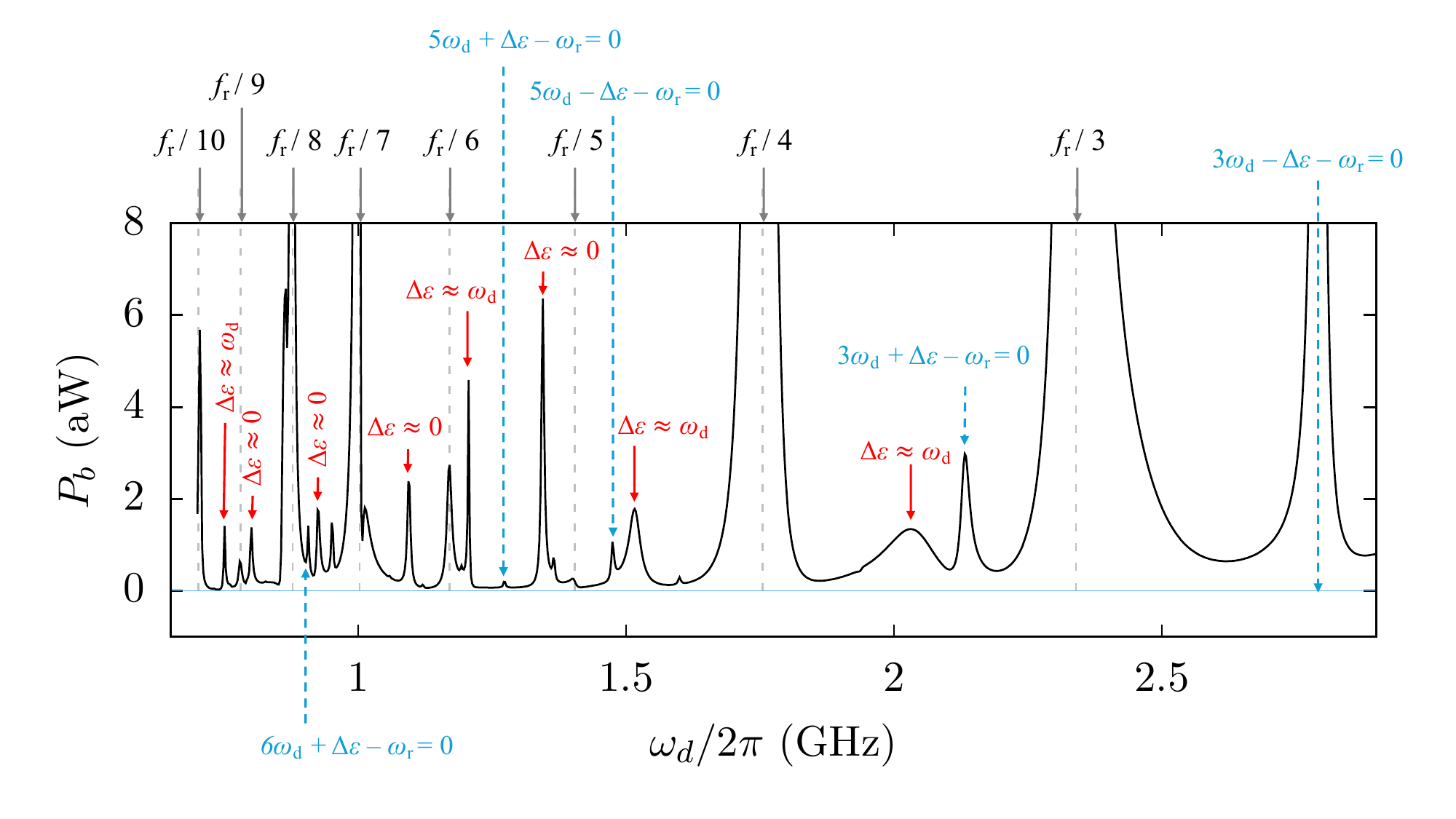}
\caption{Power to the bath \emph{vs.} drive frequency. Upper panel: Simulation of a two-level system interacting with an effective bath. Flux bias $\Phi_{\text{ext}}=0.52~\Phi_0$ and drive amplitude $A_{\text d}/2\pi=12$ GHz. The main peaks are at the fractional frequencies $\omega_{\text d}/2\pi=f_{\text r}/l$ with additional features present at intermediate frequencies where the condition~\eqref{condition} is satisfied within the width of the effective spectral density function $G_{\text{eff}}(\omega)$, see Eq.~\eqref{Geff}. Lower panel: detail of the region enclosed by the rectangle in the upper panel.  Other parameters are $\Delta/2\pi=4.708$ GHz, $f_{\text r}=\omega_{\text r}/2\pi=7.028$ GHz, $g/2\pi=0.42$ GHz, effective quality factor, see Eq.~\eqref{TLSparameters}, $\tilde Q=110$, bath temperature $T=130$~mK, and $\tilde\alpha=0.066~\omega_{\text r}^2$.    } 
\label{simulationTLS}
\end{center}
\end{figure}

Assume that the bath is a structured bath peaked at the frequency $\omega_{\text r}$. If the peak is sufficiently narrow, the spectral density function $G_{\rm eff}(\omega)$, Eq.~\eqref{Geff}, will filter-out the off-resonant contributions to the power to the bath, Eq.~\eqref{J}, namely only the terms satisfying 
\be{condition}
\omega_{nm}^l=\varepsilon_n -\varepsilon_m + l\omega_{\text d}\simeq\omega_{\text r}
\ee
will contribute.
The heat current as a function of the drive frequency displays:\\
\indent i) Main peaks at the fractional frequencies  $\omega_{\text d}=\omega_{\text r}/l$ which satisfies   the condition~\eqref{condition} for $n=m$. This  implies that only the diagonal matrix elements of $\hat Q$ contribute to the rates in Eq.~\eqref{J}.\\ 
\indent ii) Secondary peaks appear away from the fractional frequencies, provided that the quasienergies are nondegenerate, $\Delta\varepsilon=\varepsilon_1-\varepsilon_0\neq 0,\omega_{\text d}$. Indeed the resonance condition is also attained as 
$$\omega_{\text d}=\frac{\omega_{\text r} \pm\Delta\varepsilon}{l}\;.$$
In this case only the non-diagonal matrix elements contribute to the rates in Eq.~\eqref{J}.\\
\indent iii)
Finally, provided that the frequency is not to far from a fractional frequency, as compared to the width of the peak in the spectral density function $G_{\text{eff}}$,  additional peaks appear for quasi-degenerate quasienergies $\Delta\varepsilon\simeq 0,\omega_{\text d}$. In this case both diagonal and off-diagonal matrix elements contribute to the rates in the heat current formula.\\
\indent In Fig.~S\ref{simulationTLS} we show a simulation of the TLS-effective bath model displaying the power to the bath as a function of the drive frequency in the presence of a static applied bias. The position of the main and secondary peaks is also specified according to the above classification.\\ 
\indent 
In the following, we consider more in depth the case of fractional frequencies, $\omega_{\text d}=\omega_{\text r}/l$.

\subsubsection{Fractional frequency, $\omega_{\text d}=\omega_{\text r}/l$ and non-degenerate quasienergies}

If $l\omega_{\text d}=\omega_{\text r}$ and $\varepsilon_n\neq\varepsilon_m$ then the partial rates with $n\neq m$ become small because $\Delta\varepsilon+l\omega_{\text d} \ne \omega_{\text r}$ and the dominant contributions are
\be{Gamma_nn}
\Gamma^{(l)}_{nn}&=2\pi|Q_{nn}^{l}|^2 G_{\rm eff}(\omega_{\text r})n_{\text b}(\omega_{\text r})\\
\Gamma^{(-l)}_{nn}&=2\pi|Q_{nn}^{l}|^2 G_{\rm eff}(\omega_{\text r})[n_{\text b}(\omega_{\text r})+1]
\ee
where the symmetry~\eqref{symm} was used.

\begin{figure}[ht!]
\begin{center}
\includegraphics[width=0.7\textwidth,angle=0]{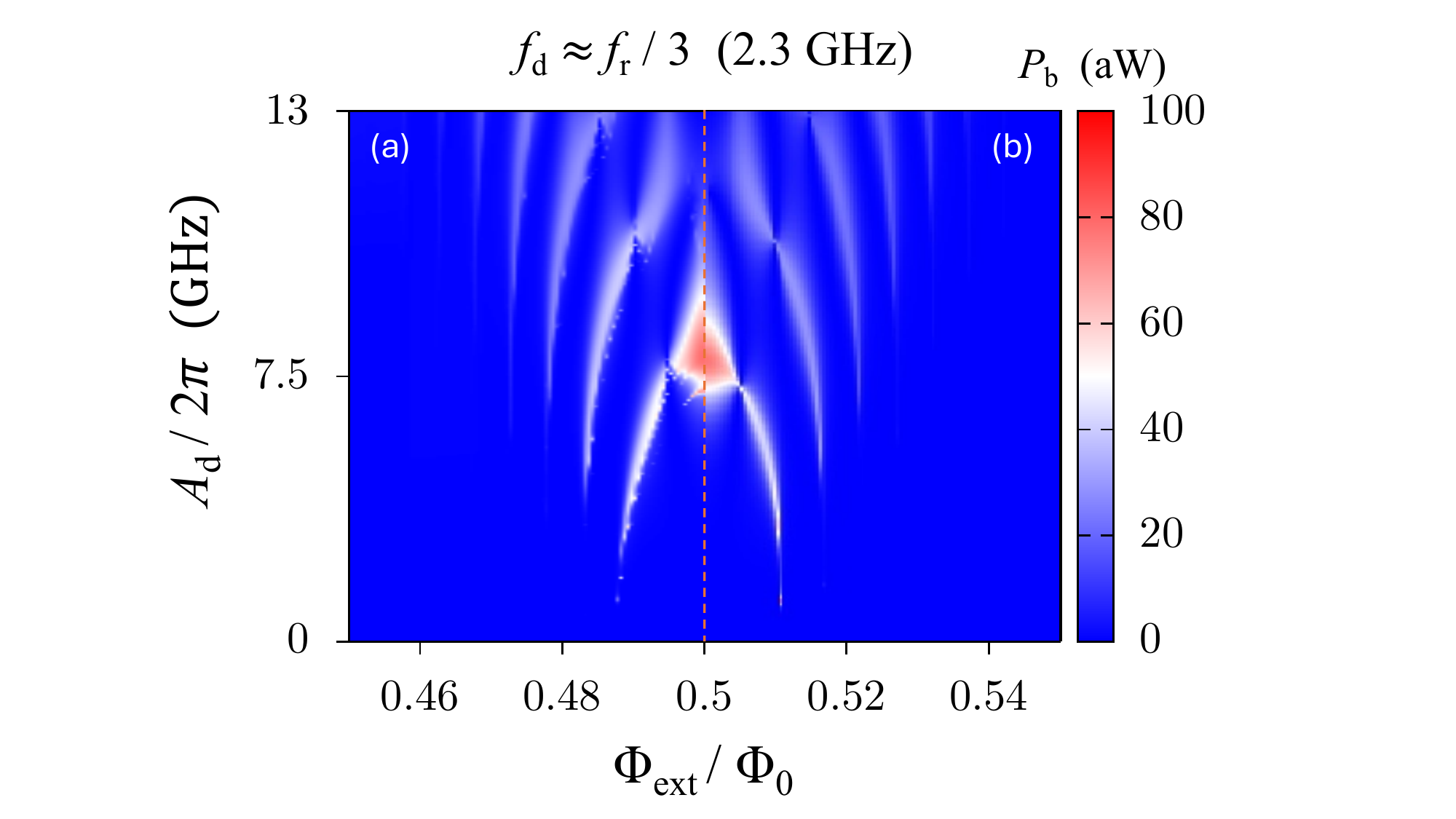}
\caption{Power to the bath \emph{vs.} drive amplitude and flux bias. The drive frequency is is close to the odd fraction $f_{\text d}\simeq f_{\text r}/3$. (a) Simulation from Eq.~\eqref{J} applied to the full quantum Rabi model with 5-level truncation of the oscillator Hilbert space. (b) Simplified formula, Eq.~\eqref{PbTLSapprox}, for the power to the bath in the effective TLS model. Other parameters are as in Fig.~S\ref{simulationTLS}.} 
\label{simulationf3}
\end{center}
\end{figure}

Still, as the diagonal elements of the rate matrix are not involved in the Pauli equation for the populations, see Eq.~\eqref{MEpopulations}, the latter are obtained solely in terms of the off-diagonal rates $m\neq n$~\footnote{Note however, that probability conservation implies $\Gamma_{nn}=-\sum_{m}\Gamma_{mn}$.}
\be{pop}
p_0=\frac{\Gamma_{01}}{\Gamma_{01}+\Gamma_{10}}
\qquad \qquad p_1=\frac{\Gamma_{10}}{\Gamma_{01}+\Gamma_{10}}\;.
\ee
\comment{
Explicitly, given that the spectral density function selects the specific value $l$ that satisfies $\pm l\omega_{\text d}=\pm\omega_{\text r}$,
the only nonvanishing contributions to the off-diagonal rates are 
\be{}
\Gamma_{01}&=\Gamma_{01}^{(l)}+\Gamma_{01}^{(-l)}\\
&=2\pi\left\{|Q_{01}^{l}|^2 G(\omega_{01}^l)n_{\text b}(\omega_{01}^l)+|Q_{01}^{-l}|^2G(\omega_{10}^l)[n_{\text b}(\omega_{10}^l)+1]\right\}\\
\Gamma_{10}&=\Gamma_{10}^{(l)}+\Gamma_{10}^{(-l)}\\
&=2\pi\left\{|Q_{10}^{l}|^2 G(\omega_{10}^l)n_{\text b}(\omega_{10}^l)+|Q_{10}^{-l}|^2 G(\omega_{10}^l)[n_{\text b}(\omega_{10}^l)+1]\right\}\;.
\ee
}
Thus, Eqs.~\eqref{J},~\eqref{definitions},~\eqref{Gamma_nn}, and~\eqref{pop} give for the dominant contribution to the heat current
\be{PbTLS}
P_{\rm b}\simeq 2\pi\hbar\omega_{\text r}G_{\rm eff}(\omega_{\text r})\left[|Q_{00}^l|^2 p_0+|Q_{11}^l|^2 p_1\right]\;.
\ee
This shows that the heat current at a fractional frequency $\omega_{\text d}=l\omega_{\text r}$, for nondegenerate quasienergies, is dictated by the diagonal matrix elements of the $l$th Fourier components of the coupling operators.
For $\hat Q=\hat \sigma_z$ (in the basis of qubit localized states) we expect $|Q_{00}^l|=|Q_{11}^l|$ yielding
\be{PbTLSapprox}
P_{\rm b}\simeq 2\pi\hbar\omega_{\text r} G_{\rm eff}(\omega_{\text r})|Q_{00}^l|^2\;.
\ee
On the basis of approximate analytical evaluations, see below, we can show that these elements vanish at the TLS symmetry point when $l$ is even. As shown in Fig. S\ref{simulationf3}, the main features of the full QRM simulations are captured  by Eq.~\eqref{PbTLSapprox}.

\subsection{Selection rule at zero static flux bias - Adiabatic limit}\label{adiabatic_limit}

In the adiabatic limit, the Floquet states of the driven TLS read $\ket{\Psi_{0/1}(t)}=\exp(\mp\ii\varepsilon_{\rm ad} t/2)\ket{\Phi_{0/1}(t)}$, where the quasienergies are given by the dynamical phase accumulated in one drive period~\cite{Koski2018}
\be{}
\varepsilon_{\rm ad}=\frac{\omega_{\text d}}{2\pi}\int^{2\pi/\omega_{\text d}}_0 dt' \omega_{\text q}(t')\;.
\ee

In the adiabatic limit the Floquet modes follow instantaneously the qubit energy levels. They read 
\be{}
\ket{\Phi_0(t)}=&\sqrt{\frac{ \omega_{\text q}(t) + \epsilon(t)}{2 \omega_{\text q}(t)}}\ket{-} - 
\sqrt{\frac{ \omega_{\text q}(t)- \epsilon(t)}{2 \omega_{\text q}(t)}}\ket{+}\\
\ket{\Phi_1(t)}=&\sqrt{\frac{ \omega_{\text q}(t)- \epsilon(t)}{2 \omega_{\text q}(t)}}\ket{-} + 
\sqrt{\frac{ \omega_{\text q}(t)+ \epsilon(t)}{2 \omega_{\text q}(t)}}\ket{+}\,,
\ee
where $\{\ket{\pm}\}$ is the qubit localized eigenbasis. In this basis, the coupling operator $\hat Q$ is $\hat\sigma_z=\ket{+}\bra{+}-\ket{-}\bra{-}$\;.
As a result the matrix elements in the Floquet basis read
\be{}
\bra{\Phi_{0/1}(t)}\hat\sigma_z\ket{\Phi_{0/1}(t)}=&\pm \frac{\epsilon(t)}{\omega_{\rm q}(t)}\\
\bra{\Phi_{0/1}(t)}\hat\sigma_z\ket{\Phi_{1/0}(t)}=&\frac{\Delta}{\omega_{\rm q}(t)}\;.
\ee
Considering the diagonal elements and expanding the denominator as 
\be{}
\frac{1}{\omega_{\rm q}(t)}=&\frac{1}{\Delta}\frac{1}{\sqrt{1+\tilde\epsilon^2(t)}}\\
=&\frac{1}{\Delta}\sum_{k=0}^\infty (-1)^k a_k[\tilde\epsilon(t)]^{2k}\;,
\ee
where $\tilde\epsilon(t)=\epsilon(t)/\Delta$ and $a_k=\prod_{s=1}^k\frac{2s-1}{2s}$, we obtain
\be{matrix_element}
\frac{\epsilon(t)}{\omega_{\rm q}(t)}=\sum_{k=0}^\infty (-1)^k a_k[\tilde\epsilon(t)]^{2k+1}\;.
\ee
Finally, in the absence of static flux bias, $\epsilon_0=0$, setting $\epsilon(t)=A_{\text d}\sin(\omega_{\text d} t)$  
\be{}
[\tilde\epsilon(t)]^{2k+1}=\frac{(A_{\text d}/\Delta)^{2k+1}}{2^{2k}}\sum_{s=k}^0 (-1)^s b_s^{(k)} \sin[(2s+1)\omega_{\text d} t]\;,
\ee
where $b_k^{(k)}=1$ and $b_{s<k}^{(k)}=[2(k-s)-1](2k+1)$. This shows that the diagonal matrix elements of the system operator $\sigma_z$ only have \emph{odd} Fourier components and therefore $Q_{nn}^l$ in the formula for the heat current, Eq.~\eqref{PbTLSapprox}, vanishes for $l$ even. Note that this symmetry is broken for $\epsilon_0\neq 0$, i.e. for nonzero static flux bias, as also even multiples of $\omega_{\text d}$ contribute to the expansion.

\section{Floquet-VVPT for a driven two-level system}

In this section we present the analytical approximate solution for the TLS Floquet quasienergies and states given by the (generalized) Van Vleck perturbation theory (VVPT)~\cite{Son2009,Hausinger2010}. The approach is based on perturbation theory in the qubit parameter $\Delta$ renormalized by the drive. It provides insight on the observed features of the heat current in the driven systems, e.g. the selection rules at the qubit symmetry point, upon evaluating the matrix elements of the system coupling operator $\hat Q$. 
As in Eq.~\eqref{HdrivenQRM}, we denote the qubit states in the localized basis with $\ket{\pm}$. In this basis $\hat Q=\hat \sigma_z$. 
 The starting point is the exact solution for $\Delta=0$, which is given, in the qubit localized basis, by the Floquet modes  
\be{Floquet_modesTLS}
|\pm_{n}(t)\rangle
&=e^{-\ii n \omega_{\text d} t}e^{\pm\ii \frac{A_{\text d}}{2\omega_{\text d}}\sin(\omega_{\text d} t)}\ket{\pm}\\
&=e^{-\ii n \omega_{\text d} t}\sum_{k=-\infty}^\infty  e^{\ii k\omega_{\text d} t}J_k\left(\pm\frac{A_{\text d}}{2\omega_{\text d}}\right)\ket{\pm}\\
&=e^{-\ii n \omega_{\text d} t}\sum_{k=-\infty}^\infty  e^{\ii k\omega_{\text d} t}\ket{\pm^{(k)}}\\
&=\sum_{k=-\infty}^\infty  e^{\ii k\omega_{\text d} t}\ket{\pm^{(k+n)}}
\;.
\ee
Note that if $\ket{u_\alpha(t)}$ is an eigenstate of the Floquet Hamiltonian, then also $\ket{u_\alpha^n(t)}=\exp(-\ii n \omega_{\text d} t)\ket{u_\alpha(t)}$ is a solution, with shifted quasi-energy. So the general form of the Floquet modes includes the reference index $n$ that shifts the quasi-energies.
The corresponding shifted quasi-energies 
are
\begin{equation}
\begin{aligned}\label{}
\varepsilon_{\pm,n}^{0}
&=\;\mp\epsilon_0/2 -n \omega_{\text d}\;.
\end{aligned}    
\end{equation}

At $\epsilon_0=0$ they form degenerate doublets, whereas at finite bias the (quasi-)degenerate doublets are the ones with 
\be{QES}
\varepsilon_{-,m}^{0}-\varepsilon_{+,0}^{0}=\epsilon_0 -m \omega_{\text d}\;,
\ee
with $m$ chosen such as to minimize the absolute value of the above difference. Accounting perturbatively for $\tilde\Delta$, the qubit tunnel splitting $\Delta$ renormalized by the drive via the Bessel functions, the (quasi-)degeneracies are lifted by the renormalized TLS tunneling splittings 
\[
\Delta_{n-l}=\Delta\langle\bra {+_{n}(t)} \hat \sigma_x\ket{-_{l}(t)}\rangle =\Delta J_{n-l}(A_{\text d}/\omega_{\text d})\;,
\]
where the double bracket indicates that the average over one driving period has been taken. 
The VVPT quasienergies read~\cite{Hausinger2010}
\be{}
\varepsilon_{\mp,n}=-n \omega_{\text d}-\frac{1}{2}m\omega_{\text d} \mp \frac{1}{2}\Omega_m \;,
\ee
where $m$ is chosen so as to minimize the quasienergy splitting, Eq.~\eqref{QES}, of the unperturbed solution. For finite $\Delta$, the splitting is given by
\be{gaps}
\Omega_m =\sqrt{\left(-\epsilon_0+m\omega_{\text d}-\frac{1}{2}\sum_{j\neq-m}\frac{|\Delta_j|^2}{\epsilon+j\omega_{\text d}} \right)^2 +\Delta_{-m}^2}\;,
\ee
and and the corresponding eigenstates, to the lowest nontrivial order in $\tilde\Delta$, are

\be{phi_pm_n}
\ket{\Phi_{\mp,n}(t)}=&\mp C_m^{\mp}\ket{+_{n}(t)}
-\tilde C_m^\pm\ket{-_{n+m}(t)}\\
&+\frac{1}{2}\sum_{j\neq-m}\frac{\Delta_j}{\epsilon_0+j\omega_{\text d}}
\left(\tilde C_m^\pm\ket{+_{j+n+m(1\pm 1)/2 }(t)} \mp C_m^\mp \ket{-_{n-j-m(1\mp 1)/2}(t)}
\right)\;.
\ee
Here 
\be{}
C^+_m&=\cos(\Theta_m/2)\;,\qquad
C^-_m=\sin(\Theta_m/2)\\
\tilde{C}_m^\pm &= {\rm sign}(\Delta_{-m})C_m^\pm\;.
\ee
The mixing angle depends on the qubit bias and reads
\be{Theta}
\Theta_m=\arctan\left(\frac{|\Delta_{-m}|}{-\epsilon_0+m\omega_{\text d}-\frac{1}{2}\sum_{j\neq-m}\frac{\Delta_j^2}{\epsilon_0+j\omega_{\text d}}}\right)\;.
\ee
From Eq.~\eqref{gaps} we note that the condition for an avoided crossing of the quasienergies are $\epsilon_0=m\omega_{\text d}$ (where $m$ is dictated by the magnitude of $\epsilon_0$ itself) and $\Delta_m=0$ (zero of the $m$-th Bessel function).\\
\indent A preliminary result, useful to calculate the matrix elements of a generic TLS coupling operator $\hat Q$ between Floquet states, is given by using Eq.~\eqref{Floquet_modesTLS} ($s=\pm$ or $\pm1$, according to the context)
\be{preliminary_result}
\sum_k \langle s^{(k)}|\hat Q| {s'}^{(k+l)}\rangle&=Q_{ss'}\sum_k J_{k}(sA_{\text d}/2\omega_{\text d})J_{k+l}(s'A_{\text d}/2\omega_{\text d})\\
&=Q_{ss'}
J_{l}[(s'-s)A_{\text d}/2\omega_{\text d}]\\
&=Q_{ss}\delta_{s' s}\delta_{l,0} + Q_{s-s}
J_{-l}(sA_{\text d}/\omega_{\text d})\delta_{s',-s}\;,
\ee
where we used the properties of the Bessel functions.
Using the above expression we find the following symmetry for the Fourier components of the matrix elements of $Q$
\be{}
\sum_k \langle s^{(k)}|\hat Q| {s'}^{(k-l)}\rangle&=Q_{ss}\delta_{s' s}\delta_{l,0} + Q_{s-s}
J_{l}(sA_{\text d}/\omega_{\text d})\delta_{s',-s}\\
&=Q_{ss}\delta_{s' s}\delta_{l,0} + Q_{s,-s}
J_{-l}(-sA_{\text d}/\omega_{\text d})\delta_{s',-s}\\
&=Q_{ss}\delta_{s' s}\delta_{l,0} + Q_{s,-s}
J_{-l}(s'A_{\text d}/\omega_{\text d})\delta_{s',-s}\\
&=\sum_k \langle {s'}^{(k)}|\hat Q| s^{(k+l)}\rangle
\ee
which verifies the symmetry of Eq.~\eqref{symm}.\\
\subsection{Matrix elements of the coupling operator in the basis of Floquet modes}
\indent Using Eq.~\eqref{phi_pm_n} we can calculate the relevant matrix elements of the coupling operator $\hat Q$ in the basis of Floquet modes. Denoting with $\alpha=\pm, n$ the composite index associated to the qubit quasienergies, we have, to the first nontrivial order in $\tilde\Delta$,
\be{}
Q_{\alpha \alpha'}^l& = \sum_k \langle \Phi_{s,n}^{(k)}|\hat Q| \Phi_{s',n'}^{(k+l)}\rangle\\
&=
\sum_k \Big(
s s' C_m^s C_m^{s'}\langle +^{(k+n)}|\hat Q| +^{(k+l+n')}\rangle - 
s C_m^s \tilde C_m^{-s'} \langle +^{(k+n)}|\hat Q| -^{(k+l+n'+m)}\rangle\\
&\qquad -s' C_m^{s'} \tilde C_m^{-s} \langle -^{(k+n+m)}|\hat Q| +^{(k+l+n')}\rangle
+\tilde C_m^{-s} \tilde C_m^{-s'}\langle -^{(k+n+m)}|\hat Q| -^{(k+l+n'+m)}\rangle
\Big)\\
&+\frac{1}{2}\sum_{j\neq-m}\frac{\Delta_j}{\epsilon_0+j\omega_{\text d}}\sum_k\Bigg[\\
&\qquad
s  C_m^s \tilde C_m^{-s'}\left(
\langle +^{(k+n)}|\hat Q| +^{(k+l+j+n'+m(1-s')/2)}\rangle
-
\langle -^{(k+n-j-m(1+s)/2)}|\hat Q| -^{(k+l+n'+m)}\rangle
\right)\\
&+ 
s s'  C_m^s C_m^{s'}\left(
\langle +^{(k+n)}|\hat Q| -^{(k+l+n'-j-m(1+s')/2)}\rangle
+
\langle -^{(k+n-j-m(1+s)/2)}|\hat Q| +^{(k+l+n')}\rangle
\right)\\
& 
-\tilde C_m^{-s} \tilde C_m^{-s'}\left(
\langle -^{(k+n+m)}|\hat Q| +^{(k+l+j+n'+m(1-s')/2)}\rangle
+
\langle +^{(k+j+n+m(1-s)/2)}|\hat Q| -^{(k+l+n'+m)}\rangle
\right)\\
& 
-s'\tilde C_m^{-s} C_m^{s'}\left(
\langle -^{(k+n+m)}|\hat Q| -^{(k+l+n'-j-m(1+s')/2)}\rangle
-
\langle +^{(k+j+n+m(1-s)/2)}|\hat Q| +^{(k+l+n')}\rangle
\right)
\Bigg]+O(\tilde\Delta^2)\;.
\ee

\subsection{Coupling operator $\hat Q=\hat\sigma_z$, in the qubit local basis}

Using the preliminary result~\eqref{preliminary_result}, the above expression specializes in the case $\hat Q=\hat\sigma_z$, to ($\alpha\equiv s,n$)
\be{MElement}
Q_{\alpha \alpha'}^l=&\sum_k \langle \Phi_{s,n}^{(k)}|\hat\sigma_z| \Phi_{s',n'}^{(k+l)}\rangle\\
=&\left(s s' C_m^s C_m^{s'} 
-\tilde C_m^{-s} \tilde C_m^{-s'}\right)\delta_{l,n-n'}\\
&+\frac{1}{2}\sum_{j\neq-m}\frac{\Delta_j}{\epsilon_0+j\omega_{\text d}}
\Bigg[
s  C_m^s \tilde C_m^{-s'}\delta_{l,-j+(n-n')-m(1-s')/2}
+s'  C_m^{s'} \tilde C_m^{-s}\delta_{l,j+(n-n')+m(1-s)/2}
\\
& 
\qquad\qquad\qquad\qquad + s C_m^{s}\tilde C_m^{-s'}\delta_{l,-j+(n-n')-m(3+s)/2}
+
s'  C_m^{s'} \tilde C_m^{-s}\delta_{l,j+(n-n')+m(3+s')/2}
\Bigg]\,.
\ee

From this result we see that it is sufficient to consider the matrix element in the so-called first Brillouin zone ($\alpha,\alpha'=-,0$ and $+,0$) and introduce the index $l'= l+n'-n$, namely $Q_{\alpha\alpha'}^l=Q_{s,n;s'n'}^l=Q_{s,0;s'0}^{l'}$

Renaming $l'\rightarrow l$, the diagonal matrix elements of $\hat Q=\hat \sigma_z$, read
\be{}
Q_{\nicefrac{-,0;-,0}{+,0;+,0}}^l
=&\quad [(C_m^\mp)^2 - (C_m^\pm)^2]  \delta_{l,0}
\\
&\mp \frac{1}{2}C_m^\mp \tilde C_m^\pm \sum_{j\neq-m}\frac{\Delta_j}{\epsilon_0+j\omega_{\text d}}
 \Bigg[ \delta_{l,-j-m(1\pm 1)/2} +\delta_{l,-j-m(3\mp 1)/2}
+ \delta_{l,j+m(3\mp 1)/2}
+ \delta_{l,j+m(1\pm 1)/2}
\Bigg]\\
\ee

\subsubsection{Special values of bias $\epsilon_0=m\omega_{\text d}$}

We relabel $j'=j+m$ which gives
\be{}
Q_{\nicefrac{-,0;-,0}{+,0;+,0}}^l
=&\quad [(C_m^\mp)^2 - (C_m^\pm)^2]  \delta_{l,0}
\\
\mp \frac{1}{2}C_m^\mp \tilde C_m^\pm &\sum_{j'\neq 0}\frac{\Delta_{j'-m}}{j'\omega_{\text d}}
 \Bigg[ \delta_{l,-j'+m-m(1\pm 1)/2} +\delta_{l,-j'+m-m(3\mp 1)/2}
+ \delta_{l,j'-m+m(3\mp 1)/2}
+ \delta_{l,j'-m+m(1\pm 1)/2}
\Bigg]\\
\ee

Explicitly

\be{diag_elements}
Q_{-,0;-,0}^l
=&\quad [(C_m^-)^2 - (C_m^+)^2]  \delta_{l,0}
- C_m^- \tilde C_m^+ 
 \Bigg[\frac{\Delta_{-l-m}}{-l\omega_{\text d}} + \frac{\Delta_{l-m}}{l\omega_{\text d}}
\Bigg]\\
Q_{+,0;+,0}^l
=&\quad [(C_m^+)^2 - (C_m^-)^2]  \delta_{l,0}
+ \frac{1}{2}C_m^+ \tilde C_m^- 
 \Bigg[\frac{\Delta_{-l}}{(-l+m)\omega_{\text d}} + \frac{\Delta_{-l-2m}}{(-l-m)\omega_{\text d}} + \frac{\Delta_{l-2m}}{(l-m)\omega_{\text d}}+ \frac{\Delta_{l}}{(l+m)\omega_{\text d}}
\Bigg]
\ee
We start by observing that, at the nodes of the heat map for $\omega_{\text d}=\omega_{\text r}/3$, the qusienergies are degenerate. In this approximate treatment, degeneracy at these special points is achieved for $\Delta_m=0$ namely for $J_m(A_{\text d}/\omega_{\text d})=0$, see the comment below Eq.~\eqref{Theta}. This, in turn implies $C^+_m=\tilde C^+_m=1$ and $C^-_m=\tilde C^-_m=0$, Eq.~\eqref{Theta}, so that the diagonal elements vanish when the two conditions $\epsilon_0=m\omega_{\text d}$ and $\varepsilon_1\simeq\varepsilon_0$ are met.

\subsubsection{Zero qubit bias: Selection rules}

In this case $m=0$ and  $C_m^\pm=\tilde C_m^\pm=\pm 1/\sqrt{2}$ which, according to Eq.~\eqref{MElement}, results in 
\be{}
Q_{s,0;s'0}^l
=&\;\frac{1}{2}(1-s s')\delta_{l,0}
+ \frac{1}{4}\sum_{j\neq0}\frac{\Delta_j}{j\omega_{\text d}}
 \Big[-2s' \delta_{l,-j}
- 2 s\delta_{l,j}
\Big]\\
=&\;\frac{1}{2}(1-s s')\delta_{l,0}
+ \frac{1}{2}\frac{\Delta_l}{l\omega_{\text d}}
 \Big[(-1)^l s' - s
 \Big]\\
=& \begin{cases}
1 \qquad\qquad \qquad \qquad\qquad s'=- s \quad l=0\\
 - s\frac{1}{2}\frac{\Delta_l}{l\omega_{\text d}}
 \Big[(-1)^l+1\Big]\qquad s'=-s\quad l\neq 0\\
 s\frac{1}{2}\frac{\Delta_l}{l\omega_{\text d}}
 \Big[(-1)^l-1\Big]\qquad  \quad s'=s \quad l\neq 0\\
\end{cases}\;.
\ee
In matrix form, the coupling operator at the qubit symmetry point, $\epsilon_0=0$, reads
\be{}
Q^{l=0}=& \begin{pmatrix}
0&1\\
1&0
\end{pmatrix}\;, \qquad 
Q^l_{(l\;\text{even})}= \frac{\Delta_l}{l\omega_{\text d}}
\begin{pmatrix}
0&-1\\
1&0
\end{pmatrix}\;,
 \qquad 
Q^l_{(l\;\text{odd})}=
\frac{\Delta_l}{l\omega_{\text d}}
\begin{pmatrix}
1&0\\
0&-1
\end{pmatrix}\;.
\ee
This shows that, for $n=n'$ the matrix element is different from zero for $l$ odd and vice-versa, for $n\neq n'$ the matrix elements do not vanish in principle for $l$ even or zero. Note that the case $l=0$ reflects the fact that the $\hat\sigma_z$ operator reads $\hat\sigma_x$ in the energy basis at zero bias. Indeed, in the static case the $l=0$ is the only nonvanishing contribution of the Fourier series. 
The so-called visibility of the qubit at zero bias follows the pattern $J_l(A_{\text d}/\omega_{\text d})$.
The selection rules emerging from this approximate picture render the suppression of the peaks of the heat flux as a function of the driving frequency when $\omega_{\rm d}=\omega_{\rm r}/l$ is an \emph{even} fraction of the resonator frequency.
Finally, note that this expression is consistent with the one given for the diagonal elements and $\epsilon_0=m\omega_{\text d}$ in Eq.~\eqref{diag_elements} above.

\subsubsection{Special points: Zeros of the $m$-th Bessel function}

Consider a bias value close to the avoided quasienergy crossing $\epsilon_0=m\omega_{\text d}$, with $m\neq 0$.
In this case, if $\Delta_m=0$ we have $C^+_m=\tilde C^+_m=1$ and $C^-_m=\tilde C^-_m=0$. This leads to 
\be{}
Q_{s,0;s'0}^l
=&\left(\delta_{s,+} 
-\delta_{s,-}\right)\delta_{l,0}\delta_{s,s'}\\
&+\frac{1}{2}\sum_{j\neq-m}\frac{\Delta_j}{\epsilon_0+j\omega_{\text d}}
\Bigg[
\delta_{s,+} \left(\delta_{l,-j-m}
+\delta_{l,-j-2m}
\right)
+\delta_{s,-} \left(\delta_{l,j+2m}
+
\delta_{l,j+m}
\right)
\Bigg]_{l\neq 0}\delta_{s,-s'} \;,
\ee
or, in matrix form
\be{}
Q^l=&\
\begin{pmatrix}
1&0\\
0&-1
\end{pmatrix}\delta_{l,0}
+
\frac{1}{2}\sum_{j\neq-m}\frac{\Delta_j}{\epsilon_0+j\omega_{\text d}}
\begin{pmatrix}
0&\delta_{l,-j-m}
+\delta_{l,-j-2m}\\
\delta_{l,j+2m}
+
\delta_{l,j+m}&0
\end{pmatrix}_{l\neq 0}\;.
\ee
Note that since for $l\neq 0$, the diagonal elements  of $Q^l$ vanish, for $l\neq 0$ this expression allows transport only for $\varepsilon_1=\varepsilon_0$, namely at the degeneracy points of the quasienergies because $\omega_{nn'}^l=l\omega_{\text r}$ even if $n\neq n'$. Note that, according to Eq.~\eqref{gaps}, $\Delta_m=0$ is also a condition for closing the gap $\Omega_m$ between quasienergies, which occurs at the bias values $\epsilon_0=m\omega_{\text d}$.

\newpage

\section{Origin of the selection rule in the adiabatic regime}

Here we give a basic picture of a driven qubit coupled to a heat bath. We analyze the case in the time domain and in adiabatic regime to describe the origin of the fractional frequencies in heat transport and their selection rules.

The master equation for a driven linear system, as in \cite{Satrya2025}, in its ground (g) and excited (e) states is given by
	\begin{eqnarray}\label{ME-1}
		&&\dot{\rho}_{\rm gg}=-\frac{2\,\lambda (t)}{\hbar}\mathfrak{Im}({\rho}_{\rm ge}e^{i\omega_0 t})-\Gamma_\Sigma {\rho}_{\rm gg}+\Gamma_\downarrow\nonumber\\&&\dot{\rho}_{\rm ge}=\frac{i\lambda (t)}{\hbar}e^{-i\omega_0 t}(2{\rho}_{\rm gg}-1)-\frac{\Gamma_\Sigma}{2}{\rho}_{\rm ge},
	\end{eqnarray}
	where $\lambda (t)=A\sin (\Omega t)$ presents the driving term, $\omega_0$ is the energy spacing of the system, $\Gamma_\downarrow,\Gamma_\uparrow$ are the relaxation and excitation rates of it, respectively, and $\Gamma_\Sigma =\Gamma_\downarrow+\Gamma_\uparrow$. 
	We need to find ${\rho}_{\rm gg}$ and for this purpose we write ${\rho}_{\rm ge}$ in the form
	\begin{equation}\label{rho_ge}
		{\rho}_{\rm ge}\equiv X+i Y.
	\end{equation}
	Using this expression, we can then rewrite the equations presented in \eqref{ME-1} as
	\begin{equation}\label{ME-rhogg2}
	\dot{\rho}_{\rm gg}=-\frac{2A}{\hbar}\sin (\Omega t) \big{[} X \sin (\omega_0 t)+ Y \cos  (\omega_0 t)\big{]} -\Gamma_\Sigma {\rho}_{\rm gg}+\Gamma_\downarrow
	\end{equation}
	\begin{equation}\label{ME-rhoge}
	\dot{X}+i \dot{Y}=\frac{iA}{\hbar} \sin (\Omega t)\big{[} \cos (\omega_0 t)-i\sin  (\omega_0 t)\big{]}(2{\rho}_{\rm gg}-1)-\frac{\Gamma_\Sigma}{2}(X+iY).
	\end{equation}
	By solving for the coherences, ignoring the initial transient and assuming a steady-state constant value of $\rho_{\rm gg}$ we have
	\begin{eqnarray}\label{XY-answers}
&&X(t)=\frac{A}{\hbar}(2{\rho}_{\rm gg}-1)\frac{\Gamma_\Sigma\cos((\Omega-\omega_0)t)+2(\Omega-\omega_0)\sin((\Omega-\omega_0)t)}{\Gamma_\Sigma^2+4(\Omega-\omega_0)^2}\nonumber\\&&Y(t)=\frac{A}{\hbar}(2{\rho}_{\rm gg}-1)\frac{-2(\Omega-\omega_0)\cos((\Omega-\omega_0)t)+\Gamma_\Sigma\sin((\Omega-\omega_0)t)}{\Gamma_\Sigma^2+4(\Omega-\omega_0)^2}.
	\end{eqnarray}
Substituting the results above in Eq.~\eqref{ME-rhogg2}, the steady-state population of the ground state ${\rho}_{\rm gg}$ in the weak coupling regime is given by
	\begin{equation}\label{rho_gg}
		{\rho}_{\rm gg}=\frac{\Gamma_\downarrow}{\Gamma_\Sigma}-\big{(}\frac{A}{\hbar\Gamma_\Sigma}\big{)}^2\frac{1}{{1+\big{(}\frac{2}{\Gamma_\Sigma}\big{)}^2(\Omega-\omega_0)^2}}.
	\end{equation}
This result applies to the case of a linear system (harmonic oscillator, resonator), but for a qubit one needs to consider the master equation 	
\begin{eqnarray} \label{f3}
	&&\dot \rho_{gg}= -\frac{\dot \epsilon}{1+\epsilon^2} \mathfrak{Im}{(e^{i\phi} \rho_{ge})}-\Gamma_\Sigma \rho_{gg}+\Gamma_\downarrow\nonumber\\&& 
	\dot \rho_{ge}=  i\frac{\dot \epsilon}{1+\epsilon^2} e^{-i\phi} (\rho_{gg}-\frac{1}{2}) -\frac{1}{2}\Gamma_\Sigma \rho_{ge}.
\end{eqnarray}
Here the drive is $\epsilon(t)=\epsilon_0+ A_{\text d}\cos (\omega_{\text d} t)$, where $ \epsilon_0$ is the bias around the symmetry point, $ A_{\text d}$ is the amplitude, and $\omega_{\text d}$ is the driving angular frequency. The phase is $\phi(t)=\omega_{\text{ge}}t$, where $\omega_{\text{ge}}/2\pi$ is the frequency of the qubit. We write 
\begin{equation}\label{lambda(t)}
\lambda(t)=\frac{\dot{\epsilon}}{1+\epsilon^2}=\dot{\epsilon}\sum_{k=0}^{\infty}(-\epsilon^2)^k,
\end{equation} which yields the harmonics besides the basic $\omega_d$. For the symmetry point $\epsilon_0=0$, we have 
\begin{equation}\label{lambda(t)expansion}
\lambda(t)=\omega_{\text d}A_{\text d}\sin(\omega_{\text d} t)\sum_{k=0}^{\infty}(-1)^kA_{\text d}^k \cos^{2k}(\omega_{\text d}t),
\end{equation}
which presents harmonics $\sin((2k+1)\omega_{\text d}t)$ with descending amplitude. If $\epsilon_0\neq 0$, we have all the odd and even harmonics $\sin(\ell \omega_{\text d}t)$, $\ell=1,\,2,\,3,..$ present. Now the solution for $\rho_{\rm gg}$ is unity except at the resonances. It is then the sum of Lorentzians of the linear equation \eqref{rho_gg}, with peaks at odd harmonics for $\epsilon_0=0$, $(2k+1)\omega_{\text d}=\omega_0$, i.e. $\omega_{\text d}=\frac{\omega_0}{2k+1}$, $k=0,\,1,\,2,…$, whereas for $\epsilon_0\neq 0$ we have the Lorentzians at all fractional frequencies $\omega_{\text d}=\omega_0/\ell$, $\ell=1,\,2,\,3,...$. This is the origin of the selection rules presented in the earlier sections. The power can be obtained trivially as 
\begin{equation}\label{power}
	P=\hbar\omega_0(\Gamma_\downarrow -\Gamma_\Sigma \rho_{\rm gg}),
\end{equation}
presenting the same Lorentzian peaks at the fractional frequencies.

\newpage

\appendix

\section{Floquet-Redfield approach to driven open quantum systems and extension to heat transport}
\label{Floquet-Redfield}

In this section, we address the heat transport problem in the driven system with the Floquet-Born-Markov master equation approach~\cite{Kohler1997,Grifoni1998}.
Consider the Born master equation~\cite{Magazzu2024PRB} 
\be{BornME}
\dot\rho(t)=&\;\mathcal{L}_S(t)\hat\rho(t) + \int_0^t dt' {\rm Tr}_B\{\mathcal{P}\mathcal{L}_{SB} \mathcal{G}_{\mathcal{Q}}^{(0)}(t,t')\mathcal{L}_{SB}\rho(t')\otimes\rho_B\}\\
=&-\frac{\rm i}{\hbar}[\tilde{H}_{\rm S}(t), \rho(t)]\\
&-\frac{1}{\hbar^2}\int_{0}^t d\tau\;\Bigg\{\left[\hat{Q}U_{\rm S}(t,t-\tau)\hat{Q}\rho(t-\tau)U_{\rm S}^\dag(t,t-\tau)-U_{\rm S}(t,t-\tau)\hat{Q}\rho(t-\tau)U_{\rm S}^\dag(t,t-\tau)\hat{Q}\right]C(\tau)\\
&\;\qquad\qquad+\left[U_{\rm S}(t,t-\tau)\rho(t-\tau)\hat{Q}U_{\rm S}^\dag(t,t-\tau)\hat{Q}-\hat{Q}U_{\rm S}(t,t-\tau)\rho(t-\tau)\hat{Q}U_{\rm S}^\dag(t,t-\tau)\right]C^*(\tau)\Bigg\}\\
\equiv & -\frac{\rm i}{\hbar}[\tilde{H}_{\rm S}(t), \rho(t)]+\int_{0}^t d\tau \;\mathcal{K}^{(2)}(t,t-\tau)\rho(t-\tau)
\;,
\ee
where $\mathcal{P}\bullet={\rm Tr}_B[\bullet]\otimes\rho_B$, with $\rho_B=\exp(-\beta\hat H_B)/Z$ the thermal state of the bath. The Liouvillian superoperators are defined as $\mathcal{L}_i \bullet=-(\ii/\hbar)[\hat H_i,\bullet]$ and the propagator is $\mathcal{G}_{\mathcal{Q}}^{(0)}(t,t')\bullet=U_{\rm S}(t,t')U_{\rm B}(t-t')\bullet U_{\rm S}^\dag(t,t')U_{\rm B}^\dag(t-t')$ (the superscript 0 stands for zeroth order in the system-bath coupling). The explicit expression for the system evolution operator is
\be{}
U_{\rm S}(t,t')=\mathcal{T}e^{-\frac{\rm i}{\hbar}\int_{t'}^t d\tau\;\hat H_{\rm S}(\tau)}\;.
\ee
Note that, in view of a second-order treatment of the system-bath interaction, in this definition we use the non-dressed Hamiltonian $\hat H_{\rm S}(t)$.
The function $C(t)$ is the  correlation function of the bath operator $\hat{B}:=\sum_j \hbar\lambda_j(\hat b_j + \hat b_j^\dag)$~\cite{Weiss2012} calculated with respect to the  thermal state $\rho_B$ of the bath 
\be{Corr_functionK}
C(t):=&\;\langle \hat B(t)\hat B(0)\rangle\\
=&\;\hbar^2\int_0^\infty d \omega\;G(\omega)\left[\coth\left(\frac{\beta\hbar\omega}{2}\right)\cos(\omega t)-{\rm i}\sin(\omega t)\right]\;.
\ee
Note that in the presence of multiple \emph{independent} baths the discussion generalizes straightforwardly: The correlation function $C(t)$ and the system operator $\hat Q$ acquire a bath index $b$ and the dissipator is summed over the baths.

\comment{
\subsubsection*{Note on the Markovin approximation}

In the interaction picture the dynamics of the reduced density matrix $\rho_I(t)$ is dictated by the system-bath interaction. Assuming that the correlation time of the function $C(t)$ is small with respect to the characteristic time of the evolution of $\rho_I(t)$, it is appropriate to set $\rho_I(t-\tau)\simeq\rho_I(t)$ (Markov approximation). Transforming back to the Schr\"odinger picture results in the Born-Markov master equation
\be{BMME}
\dot\rho(t)
=&-\frac{\rm i}{\hbar}[\tilde{H}_{\rm S}(t), \rho(t)]\\
&-\frac{1}{\hbar^2}\int_{0}^t d\tau\;\Bigg\{\left[\hat{Q}U_{\rm S}(t,t-\tau)\hat{Q}U_{\rm S}^\dag(t,t-\tau)\rho(t-\tau)-U_{\rm S}(t,t-\tau)\hat{Q}U_{\rm S}^\dag(t,t-\tau)\rho(t-\tau)\hat{Q}\right]C(\tau)\\
&\;\qquad\qquad+\left[\rho(t-\tau)U_{\rm S}(t,t-\tau) \hat{Q}U_{\rm S}^\dag(t,t-\tau)\hat{Q}-\hat{Q}\rho(t-\tau)U_{\rm S}(t,t-\tau) \hat{Q}U_{\rm S}^\dag(t,t-\tau)\right]C^*(\tau)\Bigg\}\;.
\ee
Note that this result differs from the expression one would obtain by approximating $\rho(t-\tau)\simeq\rho(t)$ in Eq.~\eqref{BornME}.
Turns out that the Markovian approximation is not necessary at the steady state, even in the presence of a time-dependent drive.}

\subsection{Floquet-Redfield master equation}

For a time-periodic system Hamiltonian with period $T=2\pi/\omega_{\text d}$, it is convenient to project the Born master equation~\eqref{BornME} in the basis of Floquet states $|\psi_n(t)\rangle$. These are the solutions of the Schr\"odinger equation for closed systems with time-periodic Hamiltonian $\tilde{H}_{\rm S}(t+T)=\tilde{H}_{\rm S}(t)$. The Floquet  theorem~\cite{Floquet1879} states that these solutions are given by
\be{Floquet_st}
|\Psi_n(t)\rangle=e^{-\ii \varepsilon_n t}|\Phi_n(t)\rangle\;,
\ee
namely 
$${\rm i}\hbar\partial_t|\Psi_n(t)\rangle = \tilde{H}_{\rm S}(t)|\Psi_n(t)\rangle\;.$$ 
The time-periodic \emph{Floquet modes} $|\Phi_n(t)\rangle=|\Phi_n(t+T)\rangle$ form a time-dependent basis of the Hilbert space of the system. Along with the quasienergies $\varepsilon_n$, they solve the eigenvalue problem
\be{QE}
[\tilde{H}_{\rm S}(t)-{\rm i}\hbar\partial_t]|\Phi_n(t)\rangle=\hbar\varepsilon_n |\Phi_n(t)\rangle\;.
\ee
We use the fact that the Floquet states solve the Schr\"odinger equation, namely 
$$U_{\rm S}(t_1,t_2)|\Psi_n(t_1)\rangle=|\Psi_n(t_2)\rangle$$
in the Born master equation~\eqref{BornME} along with  the resolution of the identity $\mathbf{1}=\sum_n |\Phi_n(t)\rangle\langle\Phi_n(t)|$ and the decomposition of the reduced density matrix in the Floquet basis 
$$\rho(t)=\sum_{n,m}\rho_{nm}(t)\ket{\Psi_n(t)}\bra{\Psi_m(t)}\;.$$ 
As a result
\be{FloquetME}
\dot \rho_{n  m }(t)&=\int_0^t d\tau \langle\Psi_n (t)|\mathcal{K}^{(2)}(t,t-\tau)\left[\sum_{ n', m'}\ket{\Psi_{ n'}(t-\tau)}\bra{\Psi_{ m'}(t-\tau)}\right]\ket{\Psi_m (t)
}\rho_{ n' m'}(t-\tau)\\    
=&-\frac{1}{\hbar^2}\sum_{ n', m'}\Bigg\{\sum_p\left[\delta_{m  m'}Q_{n p }(t)Q_{p  n'}(t-\tau)C(\tau) + 
\delta_{n  n'}Q_{ m'p }(t-\tau)Q_{p  m}(t)C^*(t)
\right]\\
&\qquad\qquad
-\left[ Q_{ m'm }(t)Q_{n  n'}(t-\tau)C(\tau) + Q_{n  n'}(t)
Q_{ m'm }(t-\tau)C^*(\tau)\right]\Bigg\}\rho_{ n' m'}(t-\tau)
\;,
\ee
where $Q_{n m }(t)=\bra{\Psi_n(t)}\hat{Q}\ket{\Psi_m (t)}$.\\ 
Note that the unitary part of the dynamics $-(\ii/\hbar)[H(t),\ket{\Psi_n(t)}\bra{\Psi_m(t)}]\rho_{n m}(t)$ cancels out with the second term of
$$\partial_t[\rho_{n m }(t)\ket{\Psi_{n }(t)}\bra{\Psi_{m }(t)}]=\dot \rho_{n m }(t)\ket{\Psi_{n }(t)}\bra{\Psi_{m }(t)} + \rho_{n m }(t)\partial_t[\ket{\Psi_{n }(t)}\bra{\Psi_{m }(t)}]\;,$$ 
see Eq.~\eqref{QE}. 

The periodicity of the Floquet modes allows for the matrix elements $Q_{n m }(t)$ to be Fourier-expanded as follows. First we expand the Floquet modes in Fourier series
\be{}
|\Phi_{n }(t)\rangle=\sum_k e^{{\rm i}k\omega t} |\Phi_n ^{(k)}\rangle\;,
\ee
then we take the matrix element of the system operator $\hat{Q}$
\be{result1}
\bra{\Psi_n (t)}\hat{Q}\ket{\Psi_m  (t)}&=\bra{\Phi_n (t)}\hat{Q}\ket{\Phi_m  (t)}e^{{\rm i }\omega_{n m  }t}\\
&=e^{{\rm i }\omega_{n m  }t}\sum_{k,k'}e^{\ii(k'-k) \omega_{\text d} t}\bra{\Phi_n ^{(k)}}\hat{Q}\ket{\Phi_m  ^{(k')}}\\
&=e^{{\rm i }\omega_{n m  }t}\sum_{l}e^{\ii l \omega_{\text d} t}\sum_{k}\bra{\Phi_n ^{(k)}}\hat{Q}\ket{\Phi_m  ^{(k+l)}}\\
&=e^{{\rm i }\omega_{n m  }t}\sum_{l}e^{\ii l \omega_{\text d} t}Q_{n m  }^{l}\;,
\ee
where $\omega_{n m  }:=\varepsilon_n  -\varepsilon_m  $. Comparing $\bra{\Phi_n  (t)}\hat{Q}\ket{\Phi_m  (t)}$ and $\bra{\Phi_m  (t)}\hat{Q}\ket{\Phi_n  (t)}^*$ one finds the symmetry 
\be{symm}
Q_{n   m  }^{-l}=Q_{m  n  }^{l}\;.
\ee
Expressing the matrix elements of $\hat Q$ in the Floquet basis leads to
\be{FloquetME}
\dot \rho_{n m}(t)  
=&-\frac{1}{\hbar^2}\sum_{n',m'}\sum_{l ,l'}e^{\ii(l-l')\omega t}\int_{0}^t d\tau\Bigg\{\sum_{p}\Big[\delta_{mm'}e^{\ii\omega_{nn'} t}Q_{np}^{-l'}Q^{l}_{p n'}e^{-{\rm i}
\omega_{p n'}^l \tau}C(\tau)\\
&\qquad\qquad\qquad\qquad\qquad\qquad\qquad+ 
\delta_{nn'}e^{\ii\omega_{m'm} t}Q_{m'p}^{-l'}Q^{l}_{p m}[e^{-{\rm i}
\omega_{p m'}^{l'} \tau}C(\tau)]^*
\Big]\\
&-e^{\ii(\omega_{nn'}-\omega_{mm'}) t}
\left[ Q_{m'm}^{-l'}Q_{nn'}^{l}e^{-{\rm i}
\omega_{nn'}^l \tau}C(\tau)
+ Q_{nn'}^{-l'}
Q_{m'm}^{l}[e^{-{\rm i}
\omega_{mm'}^{l'} \tau}C(\tau)]^*\right]\Bigg\}\rho_{n'm'}(t-\tau)
\;.
\ee
Assuming that the matrix elements in the Floquet basis of the reduced density matrix admit a time-independent stationary state $\rho_{nm}(t)\rightarrow \rho_{nm}$, one can solve the convolutive integral over $\tau$ using the final value theorem
\be{FloquetMESS}
0  
=&-\frac{1}{\hbar^2}\sum_{n  ',m'}\sum_{l ,l'}e^{\ii(l-l')\omega t}e^{\ii(\omega_{n  n'}-\omega_{mm'})t}\Bigg\{\sum_{p }\Big[\delta_{mm'}Q_{n p }^{-l'}Q^{l}_{p n'}W_{p n'}^l+ 
\delta_{nn'} Q_{m'p }^{-l'}Q^{l}_{p m}[W_{p m'}^{l'}]^*\Big]\\
&\qquad\qquad\qquad\qquad\qquad\qquad\qquad-
\left[ Q_{m'm}^{-l'}Q_{nn'}^{l}W_{nn'}^l
+ Q_{nn'}^{-l'}
Q_{m'm}^{l}[W_{mm'}^{l'} ]^*\right]\Bigg\}\rho_{n'm'}
\;,
\ee
where
\be{WalphabetaSS}
W^{l}_{nm}=&\int_{0}^\infty d\tau\; e^{-{\rm i}
\omega_{nm}^l \tau}C(\tau)\;.
\ee
Further, assuming that, at the steady sate the coherences in the Floquet basis vanish, namely $\rho_{nm}=0$ for $n\neq m$~\cite{Kohler1997}, similarly to the case of static systems in the full secular master equation approach, then Eq.~\eqref{FloquetMESS} specializes for the populations ($m=n$, $n'=m'=m$)
\be{FloquetMESSpop}
0  
=&-\frac{1}{\hbar^2}\sum_{m}\sum_{l ,l'}e^{\ii(l-l')\omega t}\Bigg\{\sum_{ p}\Big[\delta_{ n m}Q_{ n p}^{-l'}Q^{l}_{ p m}W_{ p m}^l+ 
\delta_{ n m} Q_{ m p}^{-l'}Q^{l}_{ p n}[W_{ p m}^{l'}]^*\Big]\\
&\qquad\qquad\qquad\qquad\qquad\qquad\qquad-
\left[ Q_{ m n}^{-l'}Q_{ n m}^{l}W_{ n m}^l
+ Q_{ n m}^{-l'}
Q_{ m n}^{l}[W_{ n m}^{l'} ]^*\right]\Bigg\}\rho_{ m m}
\;,
\ee
Performing the average over one time period yields $l=l'$. This in turn gives the final form of the Pauli-type master equation for the populations in the Floquet basis 
\be{}
0=\sum_ m (\Gamma_{ n m}\rho_{ m  m}-\Gamma_{ m n}\rho_{ n  n})\;,
\ee
with the rates given by 
\be{Gammal}
\Gamma_{ n m}&=\sum_l \Gamma_{ n m}^{(l)}=\frac{1}{\hbar^2}\sum_l |Q_{ n m}^l|^2 2{\rm Re} W_{ n m}^l
\;.
\ee
Here we have used the symmetry~\eqref{symm}. The Sokhotski-Plemelj theorem yields for the real part of $W$
\be{}
{\rm Re}W^{l}_{ n m}/\hbar^2=&\;\frac{\pi}{2}G(|\omega^{l}_{ n m}|)\left[\coth\left(\frac{ m\hbar|\omega^{l}_{ n m}|}{2}\right)-\frac{\omega^{l}_{ n m}}{|\omega^{l}_{ n m}|}\right]\\
=&\;
\begin{cases}
\pi G(|\omega^{l}_{ n m}|)  n_{\text b}(|\omega^{l}_{ n m}|) \qquad \qquad \omega^{l}_{ n m}\geq 0\\
\pi G(|\omega^{l}_{ n m}|) [  n_{\text b}(|\omega^{l}_{ n m}|)+1]\qquad\omega^{l}_{ n m}<0\;.
\end{cases}
\;,
\ee
If the bath spectral density, extended to negative frequencies, is an odd function of $\omega$ we can simply write
$
{\rm Re}W^{l}_{ n m}/\hbar^2=\pi G(\omega^{l}_{ n m})  n_{\text b}(\omega^{l}_{ n m})
$.

\subsection{Current}

The current operator to the bath is $\hat I_{\text b}=d\hat H_B/d t=\hat Q\otimes \bar{B}$, with  $\bar{B}:={\rm i} \sum_{j}\lambda_{j}\hbar\omega_{j}[b_j - b^\dag_j]$, while $\hat H_{\rm SB}=\hat Q\otimes \hat{B}$ with $\hat{B}:=\sum_j\hbar\lambda_{j}(\hat b_j + \hat b_j^\dag)$, see Eq.~\eqref{Hterms}.
Thus, to second order, we obtain for the heat power to the bath~\cite{Magazzu2024PRB} 
\be{}
P_{\rm b}(t)&=\int_0^t dt' {\rm Tr}\{\hat I_{\text b} \mathcal{G}_{\mathcal{Q}}^{(0)}(t,t')\mathcal{L}_{\rm SB}\rho(t')\otimes\rho_B\}\\
&=-\frac{\ii}{\hbar}\int_0^t dt'\;{\rm Tr}\Big\{\hat I_{\text b} \mathcal{G}_{\mathcal{Q}}^{(0)}(t,t')\left[H_{{\rm SB}}, \rho(t')\otimes \rho_B\right]\Big\}\\
&= -\frac{\ii}{\hbar}\int_0^t d\tau\;{\rm Tr}_{\rm S}\Big\{\hat{Q} U_{\rm S}(t,t-\tau)\hat{Q} \rho(t-\tau) U_{\rm S}^\dag(t,t-\tau) K(\tau)\\
&\qquad\qquad\qquad 
-\hat{Q} U_{\rm S}(t,t-\tau)\rho(t-\tau)\hat{Q}  U_{\rm S}^\dag(t,t-\tau) K^*(\tau)\Big\}\;,
\ee
where we introduced the correlator~\footnote{Note that, as $\bar{B}$ is Hermitian, $\langle \hat B(t')\bar{B}(t)\rangle=K^*(t-t')$.} 
\be{}
K(t)=&\;\langle \bar{B}(t)\hat B(0)\rangle\\
=&\;\ii \hbar^2\sum_j\omega_j \lambda_j^2\left[e^{-\ii\omega_j t}\langle \hat b_j \hat b_j^\dag \rangle - e^{\ii\omega_j t}\langle \hat b_j^\dag \hat b_j \rangle \right]\\
=&\;\ii\hbar^2\sum_j \omega_j \lambda_j^2\left[e^{-\ii\omega_j t}[1+n_{\text b}(\omega_j)] - e^{\ii\omega_j t} n_{\text b}(\omega_j) \right]\\
=&\;\hbar^2\sum_j \omega_j\lambda_j^2\left\{[1+2 n_{\text b}(\omega_j))]\sin(\omega_j t) +\ii \cos(\omega_j t) \right\}\\
=&\;\hbar^2\int_0^\infty d \omega\;\omega G(\omega)\left[\coth\left(\frac{\beta_{\text b}\hbar\omega}{2}\right)\sin(\omega t)+{\rm i}\cos(\omega t)\right]\\
\equiv &\;-\frac{d}{dt}C(t) \;.
\ee
Here, $C(t)$ is the bath correlation function, Eq.~\eqref{Corr_functionK}.
At this point we project in the basis of Floquet states~\eqref{Floquet_st} Performing the trace as $\sum_n \bra{\Psi_n (t)}\bullet\ket{\Psi_n (t)}$ we obtain
\be{}
P_{\rm b}(t)=&\frac{\ii}{\hbar}\sum_{n ,n',m'}\int_0^t d\tau\;\Big\{ 
Q_{m'n }(t)Q_{n n '}(t-\tau)\frac{d}{d\tau}C(\tau)
-
Q_{n n '}(t)Q_{m'n }(t-\tau)\frac{d}{d\tau}C^*(\tau)\Big\}\rho_{n 'm'}(t-\tau)\;,
\ee
Using the expansion~\eqref{result1} for the system coupling operator
\be{}
P_{\rm b}(t)=&\frac{\ii}{\hbar}\sum_{n ,n ',m'}\sum_{l,l'}e^{\ii(l-l')\omega_{\text d} t}\int_0^t d\tau\;\Big\{ 
Q_{m'n }^{-l'}Q_{n n '}^{l}e^{\ii\omega_{m'n '}t}
e^{-{\rm i}
\omega_{n n '}^l \tau}\frac{d}{d\tau}C(\tau)
\\
&\qquad\qquad\qquad-
Q_{n n '}^{-l'}Q_{m'n }^{l}e^{\ii\omega_{m'n '}t}
[e^{-{\rm i}
\omega_{n m'}^{l} \tau}\frac{d}{d\tau}C(\tau)]^*
\Big\}\rho_{n 'm'}(t-\tau)\;,
\ee
Assuming that, at the steady state, $\rho_{n m}(t)=\rho_{n m}$, we can solve the convolutive integral over $\tau$ using the final value theorem and define
\be{}
\bar W^{l}_{n\beta}=\int_{0}^\infty d\tau\; e^{-{\rm i}
\omega_{nm}^l \tau}\frac{d}{d\tau}C(\tau)&=[e^{-{\rm i}
\omega_{nm}^l \tau}C(\tau)]_0^\infty -  \int_{0}^\infty d\tau\; \frac{d}{d\tau}(e^{-{\rm i}
\omega_{nm}^l \tau})C(\tau)\\
&=-C(0) +\ii \omega_{nm}^l \int_{0}^\infty d\tau\; e^{-{\rm i}
\omega_{nm}^l \tau}C(\tau)\\
&=-C(0) +\ii \omega_{nm}^l W^{l}_{nm}\;,
\ee
where $W^{l}_{nm}$ is defined in Eq.~\eqref{WalphabetaSS}. If, as above, we asume that, at the steady state, $\rho_{nm}=0$ for $n\neq m$, the current reads
\be{}
P_{\rm b}(t)=&\frac{\ii}{\hbar}\sum_{ n m}\sum_{l,l'}e^{\ii(l-l')\omega t}\Big\{ 
Q_{ m n}^{-l'}Q_{ n m}^{l}\bar W_{ n m}^l-
Q_{ n m}^{-l'}Q_{ m n}^{l}(\bar W_{ n m}^{l})^*\Big\}\rho_{ m m}\;.
\ee
Taking the one-period average (which yields $l=l'$) 
\be{}
P_{\rm b}=&\frac{\ii}{\hbar}\sum_{ n, m,l}| Q_{ n m}^{l}|^2 2\ii{\rm Im}(\bar W_{ n m}^{l})\rho_{ m m}\\
=&-\sum_{ n, m,l}\hbar\omega_{ n m}^l\Gamma_{ n m}^{(l)}\rho_{ m m}
\;,
\ee
where $\Gamma_{ n m}^{(l)}$ is the partial rate defined in Eq.~\eqref{Gammal}.

 \bibliography{SIReferences}